\documentclass{article}
\pdfoutput=1
\usepackage{arxiv}

\usepackage[utf8]{inputenc} 
\usepackage[T1]{fontenc}    
\usepackage{hyperref}       
\usepackage{url}            
\usepackage{booktabs}       
\usepackage{amsfonts}       
\usepackage{nicefrac}       
\usepackage{microtype}      
\usepackage{graphicx}
\usepackage{natbib}
\usepackage{doi}
\usepackage{pdfpages}
\usepackage{chronology}
\usepackage{tabularx}

\title{Machine Learning of Public Sentiments toward Wind Energy in Norway}

\date{April 4, 2023}	

\author{
    \href{https://orcid.org/0000-0002-0806-0329}{\includegraphics[scale=0.06]{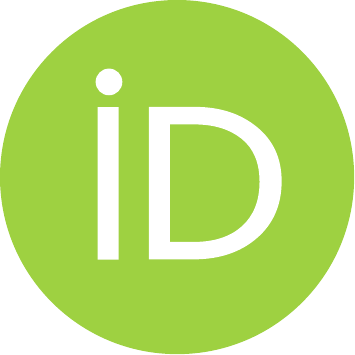}\hspace{1mm}}
	Oskar Vågerö\thanks{Corresponding Author} \\
	Department of Technology Systems\\
	University of Oslo\\
    Kjeller, Norway \\
	\texttt{oskar.vagero@its.uio.no} \\
    \And
    Anders Bråte \\
	Department of Physics \\
	University of Oslo \\
    Oslo, Norway \\
	\texttt{anders.brate@fys.uio.no} \\
	\And
    Alexandra Wittemann \\
	Department of Informatics\\
	University of Oslo\\
    Oslo, Norway\\
	\texttt{alexankw@ifi.uio.no} \\
	\And
	Jessica Yarin Robinson \\
	Department of Media and Communications \\
	University of Oslo \\
	Oslo, Norway \\
	\texttt{j.y.robinson@media.uio.no}
	\And
	Natalia Sirotko-Sibirskaya \\
	Department of Mathematics \\
	University of Oslo \\
	Oslo, Norway \\
	\texttt{nsibirska@math.uio.no} \\
	\And
    \href{https://orcid.org/0000-0003-1756-1878}{\includegraphics[scale=0.06]{orcid.pdf}\hspace{1mm}}
	Marianne Zeyringer \\
	Department of Technology Systems\\
	University of Oslo\\
	Kjeller, Norway \\
	\texttt{marianne.zeyringer@its.uio.no} \\
}

\renewcommand{\shorttitle}{Machine Learning of Public Sentiments toward Wind Energy in Norway}

\hypersetup{
pdftitle={Machine Learning of Public Sentiments toward Wind Energy in Norway},
pdfsubject={stat.AP},
pdfauthor={Oskar Vågerö, Anders Bråte, Alexandra Wittemann, Jessica Yarin Robinson, Natalia Sirotko-Sibirskaya, Marianne Zeyringer, },
pdfkeywords={Wind power, Machine learning, Sentiment analysis, Public sentiment},
colorlinks=true,
linkcolor=blue,
filecolor=magenta,
citecolor=blue,
urlcolor=cyan,
}

\usepackage{tikz}
\usepackage{mathtools}
\usepackage{graphicx}
\usetikzlibrary{positioning, shapes.geometric, shapes.symbols, arrows, shapes.misc}
\usetikzlibrary{positioning,chains}
\usetikzlibrary{snakes}
\usetikzlibrary{decorations.pathreplacing,calligraphy}

\tikzset{
    mynode/.style={
        draw, rectangle, align=center, text width=5cm, font=\small, inner sep=2ex},
    mylabel/.style={
        draw, rectangle, align=center, rounded corners, font=\small\bf, inner sep=2ex,
        fill=cyan!30, minimum height=3.8cm},
    arrow/.style={
        very thick,->,>=stealth}
}

\tikzset{
    database/.style={
        path picture={
            \draw (0, 1.5*\database@segmentheight) circle [x radius=\database@radius,y radius=\database@aspectratio*\database@radius];
            \draw (-\database@radius, 0.5*\database@segmentheight) arc [start angle=180,end angle=360,x radius=\database@radius, y radius=\database@aspectratio*\database@radius];
            \draw (-\database@radius,-0.5*\database@segmentheight) arc [start angle=180,end angle=360,x radius=\database@radius, y radius=\database@aspectratio*\database@radius];
            \draw (-\database@radius,1.5*\database@segmentheight) -- ++(0,-3*\database@segmentheight) arc [start angle=180,end angle=360,x radius=\database@radius, y radius=\database@aspectratio*\database@radius] -- ++(0,3*\database@segmentheight);
        },
        minimum width=2*\database@radius + \pgflinewidth,
        minimum height=3*\database@segmentheight + 2*\database@aspectratio*\database@radius + \pgflinewidth,
    },
    database segment height/.store in=\database@segmentheight,
    database radius/.store in=\database@radius,
    database aspect ratio/.store in=\database@aspectratio,
    database segment height=0.1cm,
    database radius=0.25cm,
    database aspect ratio=0.35,
}
\makeatother

\begin{document}
\maketitle

\begin{abstract}
	Across Europe negative public opinion has and may continue to limit the deployment of renewable energy infrastructure required for the  transition to net-zero energy systems. Understanding public sentiment and its spatio-temporal variations is as such important for decision-making and socially accepted energy systems. In this study, we apply a sentiment classification model based on a machine learning framework for natural language processing, NorBERT, on data collected from Twitter between 2006 and 2022 to analyse the case of wind power opposition in Norway. From the 68828 tweets with geospatial information, we show how discussions about wind power intensified in 2018/2019 together with a trend of more negative tweets up until 2020, both on a regional level and for Norway as a whole. Furthermore, we find weak geographical clustering in our data, indicating that discussions are country wide and not dominated by specific regional events or developments. Twitter data allows for detailed insight into the temporal nature of public sentiments and extending this research to additional case studies of technologies, countries and sources of data (e.g. newspapers, other social media) may prove important to complement traditional survey research and the understanding of public sentiment.
\end{abstract}


\newpage
\section{Introduction}
Wind power technology is considered key for the transition to net-zero energy systems \citep{nordicenergyresearchNordicEnergyTechnology2016,irenaGlobalRenewablesOutlook2020,ieaRenewables20222022,europeancommissionCommunicationCommissionEuropean2022} and to increase Europe's energy independence, yet it is also contested in many countries, including the UK \citep{roddisRoleCommunityAcceptance2018}, Denmark \citep{ladenburgOffshoreonshoreConundrumPreferences2020} and Norway \citep{normannGreenColonialismNordic2021}. Negative public sentiment toward wind energy can act as an impediment to developments that are otherwise techno-economically viable \citep{devine-wrightNIMBYismIntegratedFramework2005,johanssonIntentionRespondLocal2007,firestonePublicAcceptanceOffshore2009}. The importance of wind energy to meet climate targets as well as responding to the energy crisis in Europe thus makes it important to understand public sentiments.

The role of public acceptance of onshore wind power has been a topic of research ever since the early 1980s \citep{wustenhagenSocialAcceptanceRenewable2007} and studies have attempted to identify drivers of acceptance/opposition, such as national and local ownership and use \citep{warrenDoesCommunityOwnership2010,brennanPublicAcceptanceLargescale2017,linnerudPeoplePreferOffshore2022,vuichardKeepItLocal2022}, exposure and proximity to the wind power plants \citep{brennanWindFarmExternalities2016,dugstadAcceptanceWindPower2020} and associated ecological impacts \citep{vuichardKeepItLocal2022}. Results are not always consistent and while some indicate that for example exposure to wind power plants lead to higher acceptance \citep{liebeTurbineNotOnly2017}, others show the opposite \citep{dugstadAcceptanceWindPower2020}. Recent research have highlighted geographical difference and the need for more cross-country comparisons \citep{vuichardKeepItLocal2022}, suggesting that both temporal and spatial aspects of public acceptance need further exploration.

Survey research and interviews have traditionally been important methods to acquire data on public opinions. However, these methods have limitations which may induce a bias into an analysis, such as coverage-, sampling-, measurement- and nonresponse error \citep{pontoUnderstandingEvaluatingSurvey2015}. While there are scientifically tested strategies to reduce these shortcomings and for improving the rigour of a study \citep{pontoUnderstandingEvaluatingSurvey2015, sovacoolPromotingNoveltyRigor2018}, survey research may also be constrained by the size of the sample investigated, especially when surveys are conducted by phone, over e-mail or in-person. \citet{devine-wrightNIMBYismIntegratedFramework2005} has called for new methodologies that go beyond surveys and capture other dimensions of public perception of wind energy (p. 135).

In recent years, new developments in data science and computer models for Natural Language Processing (NLP), combined with an increasing activity on social media, have opened up new ways of acquiring and analysing large amounts of public opinion data. For example, this makes it possible to explore longer time horizons and study how public opinion has evolved over time, or different geographies and technologies. Currently, to our knowledge there is no study applying NLP to Twitter data to perform a sentiment analysis on wind energy.

Here, to close this gap, we seek to understand the public sentiments of wind energy across Norway. This Scandinavian country presents a useful case because the country has set ambitious goals for reducing greenhouse gas emissions, yet efforts to expand wind energy production have been hampered by a rise in public opposition. With this as the backdrop, we demonstrate the use of a sentiment classification model for the Norwegian language, NorBERT \citep{kutuzovLargeScaleContextualisedLanguage2021}, to data collected from Twitter between 2006 and 2022 on the public opinion towards wind power in Norway and compare our results with data from two surveys. Our findings suggest that this method, even when limited to a single platform, can be useful for tracking the temporal patterns of public attention to wind power, and help inform fluctuations in opinion surveys.

As such, we contribute to the growing literature on 1. public opinions towards renewable energy technologies and better understanding of these opinions and their spatio-temporal variation, 2. when machine learning methods such as NLP are useful to either contrast or complement traditional survey research and 3. more generally on using Twitter data to analyse questions on climate change mitigation.

The paper is structured as follows. In section 2 we describe the context in which wind energy debates are playing out in Norway. Following this, in section 3, we examine Twitter as a platform for public discussions of wind energy and previous research using NLP. Section \ref{sec:research-design} then outlines the process of data collection, pre-processing of the data, fine-tuning of NorBERT and a general overview of the final dataset. Section \ref{sec:results} presents the results of the sentiment analysis, which are further discussed in section \ref{sec:discussion}.

\section{Wind energy in Norway}
Excellent wind resources, technological development and increased interconnection to European energy markets contributed to the rapid development of a previously small wind power sector in Norway \citep{nveForslagTilNasjonal2019, inderbergWhoInfluencesWindpower2019}, where the installed capacity increased more than threefold between 2016 and 2021 \citep{vasstromWhatShapesNorwegian2021} from 1.0GW to 4.2GW corresponding to 3.0\% and 10.7\% of total installed generation capacity \footnote[1]{Hydropower capacities- 31.6GW (2016), 34.1 GW (2021), Thermal power capacities 1.1 GW (both 2016, 2021)}. Similar to other European countries, the rapid increase in new wind power development led to conflicts around e.g. its environmental impact on biodiversity, landscape aesthetics and ownership structures \citep{dugstadAcceptanceWindPower2020, linnerudPeoplePreferOffshore2022}, and when the Norwegian Water Resources and Energy Directorate (NVE) presented their proposed National Framework for land-based Wind Power (NFWP) in April 2019, it sparked new debates and protests. The public opposition against wind power developments eventually led to the withdrawal of the proposed national framework, a temporary stop in the granting of new licencing and more restrictive rules for wind power licensing \citep{norwegianministryofpetroleumandenergyMeldSt282020}.

The public attitude towards wind power has seen a significant shift in the last few years, with the support for building more onshore wind power dropping from 65\% to 33\% and opposition increasing from 10\% to 40\% between 2018 and 2021 \citep{aasenFolkOgKlima2022}.

\begin{figure}[h!]
    \centering
    \includegraphics[width=0.7\textwidth]{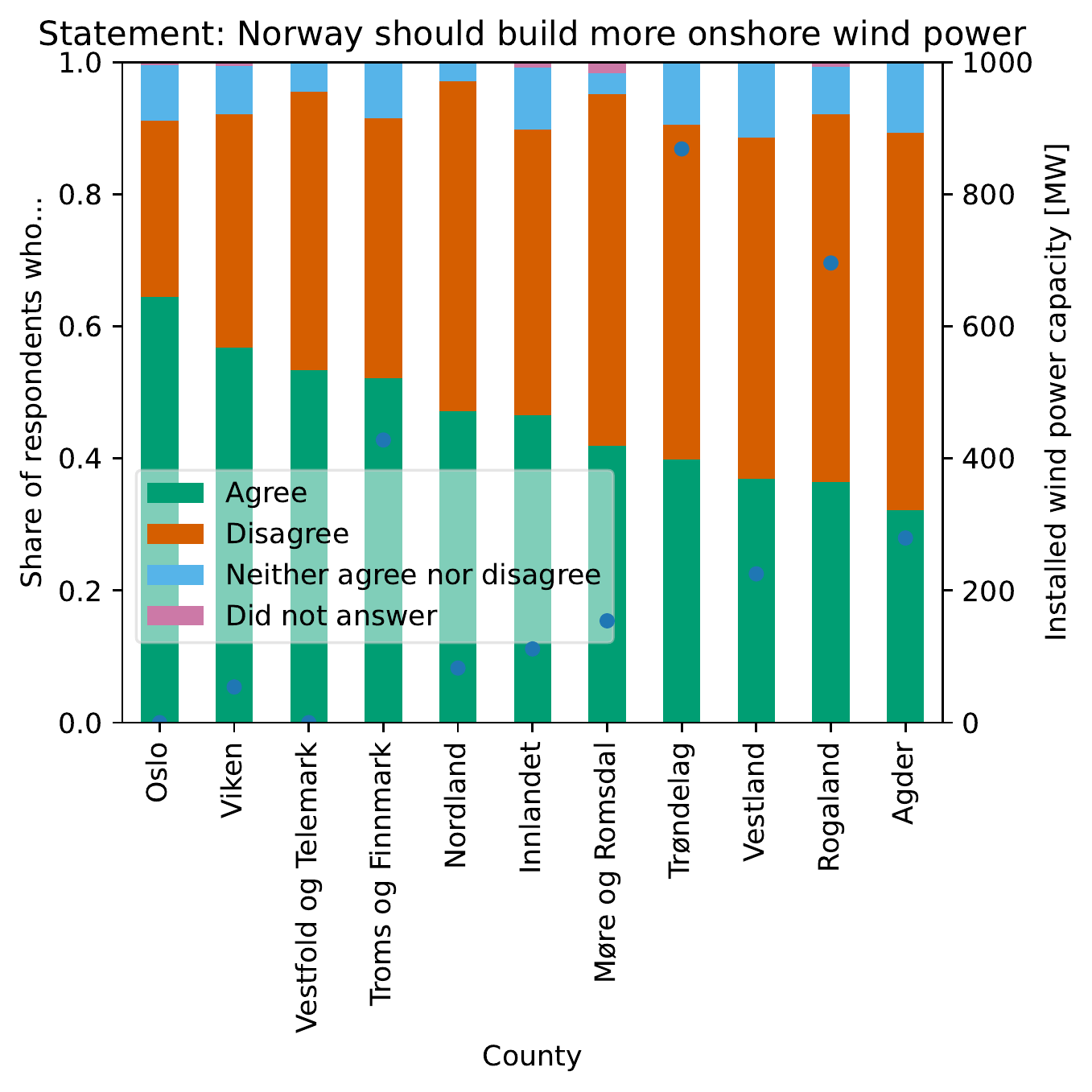}
    \caption{Spatial variation (NUTS3 level) in support for more onshore wind, based on \citet{ivarsflatenNorwegianCitizenPanel2020}} with blue dots representing the installed wind capacity in 2019 (MW)
    \label{fig:uib_onshore}
\end{figure}

Furthermore, surveys indicate, as seen in figure \ref{fig:uib_onshore}, that there are regional differences in the support for onshore wind power, with the lower support in areas with much wind power development, such as south-western Norway (Rogaland) and central-Norway (Trøndelag).


\section{Twitter as a site of public sentiment}
Recent work by \citet{kimPublicSentimentSolar2021} used Natural Language Processing (NLP) and data from the social media platform 'Twitter' to perform sentiment analysis of solar energy in the United States. NLP is a branch within computer science in which computers are built to understand, and respond to, text or voice data given to it. It may be used for things such as speech recognition or, in the case of understanding public perception, sentiment analysis \citep{ibmcloudeducationWhatNaturalLanguage2020}. Sentiment analysis is the classification of text, typically either positive (\textit{The world needs more wind energy}), negative (\textit{Wind power plants are harmful to the environment}) or neutral \citep{tonkinChapterDayWork2016}. \citet{kimPublicSentimentSolar2021} sourced data from Twitter, a social media for public communication, where users post and interact with 'tweets', i.e. shorter messages of up to 280 characters \citep{brunsNorwegianTwittersphere2018}. Compared to other platforms, messages are short, fast-paced and close to real-time communication. As a result, Twitter has come to be dominated by posts about news and public affairs, rather than personal updates \citep{burgessTwitterBiography2020}.

In Norway, as in many other countries, Twitter is a prime means of communication among journalists, politicians, heads of NGOs, and other opinion leaders \citep{larssonTriumphUnderdogsComparing2014}. As of 2022, over a quarter of Norwegian adults had a Twitter profile \citep[p. 4]{ipsosSOSIALEMEDIERTRACKER2022}, with men (33\%) more likely than women (24\%) to have a profile. While Twitter is not as widely used in the Norwegian population as Facebook or Instagram \citep{ipsosSOSIALEMEDIERTRACKER2022}, it is seen as an influential niche channel for the politically engaged. Previous research suggests Twitter may be especially important for discussions about environmental issues in Norway \citep{rogstadTwitterJustRehashing2016}, and specifically climate change \citep{kirilenkoPublicMicrobloggingClimate2014}.

Twitter is furthermore a relatively open platform, with only a small percentage of the users setting their profile as private \citep{brunsNorwegianTwittersphere2018}, which together with a well developed application program interface (API) makes the data very accessible for e.g. research purposes. Examples of studies utilising Twitter are \citet{falkenbergGrowingPolarizationClimate2022}, who uses Twitter data from 2014 to 2021 to study the discussion around the United Nations Conference of the Parties on Climate Change, \citet{codyClimateChangeSentiment2015} who perform a climate change sentiment analysis using data from 2008 to 2014. \citet{labonteTweetsTransitionsExploring2021} analyse Twitter-based discourse regarding energy issues and politics in Ontario, Canada from between September 2, 2017 and January 12, 2018 and \citet{muller-hansenGermanCoalDebate2022} use Twitter data to analyse the German coal debate.

While \citet{kimPublicSentimentSolar2021} explored spatial variations in the support towards solar energy in the United States there are other contexts and opportunities where applying NLP to large datasets from e.g. Twitter may be of interest to study sentiments towards climate change mitigation technologies and instruments  .

\section{Materials \& methods}
\label{sec:research-design}
In this study, we draw upon the methodology of \citet{kimPublicSentimentSolar2021}, but in a new context: wind power in Norway as opposed to solar energy in the United States. Information on the public sentiment towards wind power in Norway was acquired through data scraping of the social media platform Twitter via its API. The acquired data was subsequently manually pre-processed before machine learning methods for Natural Language Processing (NLP) were applied on the data set to conduct the sentiment analysis. The data collection is described in section \ref{sec:data_coll}, whilst section \ref{sec:training_model} gives an overview of NorBert and the application of the machine learning algorithm.

\subsection{Data collection}
\label{sec:data_coll}

For the collection of data, we extracted user-created posts (tweets) via the Twitter API, which was queried with a set of keywords and a time interval, to which it will return tweets matching the given keywords. Only tweets in the Norwegian language were considered relevant for the analysis. The keywords included in the API query were: '(havvind OR vindkraft OR vindmølle OR vindmøller OR vindmøllene OR vindturbiner OR vindenergi) lang:no', which translates to offshore wind, wind power, wind mill, wind mills, wind turbines, wind energy. The keywords used in the query were chosen as the Twitter API limits this number to 8 parameters, of which variations in spelling has to be explicitly stated.

In addition to the text of the tweet itself the API returned a tailored amount of additional data regarding the tweet and the user in so-called fields\footnote{\url{https://developer.twitter.com/en/docs/twitter-api/fields}}. These fields provide information including, but not limited to, the amount of likes, the amount of retweets, the full username, the tweet's geolocation if there is one, and the user's location. To build a spatially resolved dataset, which allows analysing developments in different Norwegian regions, we need information on the geography of the tweets. The tweet location is by far the most accurate description of the user location, however, as this is an opt-in feature on Twitter, meaning that it by default is turned off, its use for our data set is limited. We therefore combine this with the user defined location, given primarily as the user makes the Twitter account. The limitations this brings however is that this is simply a string input field, which is not limited to actual geographical locations. We resolved this by iterating through a list of Norwegian towns and cities sourced primarily from Wikipedia, similar to the method of \citet{brunsAustralianTwittersphere20162017}. Tweets that either had a location field in the tweet, or an intelligible location in the user field, and which was a place in Norway were kept as part of the main data set. For training NorBERT, it was not necessary for the tweet to be associated with a location, meaning that the training data could be sampled from the larger data set.

The tweets collected goes back to the first tweet related to wind energy, in 2006. Since the Norwegian Twitter landscape is relatively small in absolute numbers\footnote{as compared to e.g. the one of the U.S.} (between 716 000 and 1 120 000 users depending on the method of calculation \citep{brunsNorwegianTwittersphere2018,ipsosSOSIALEMEDIERTRACKER2022}) this is of no hindrance, and it is possible to analyse the sentiment over time periods after extracting the tweets.

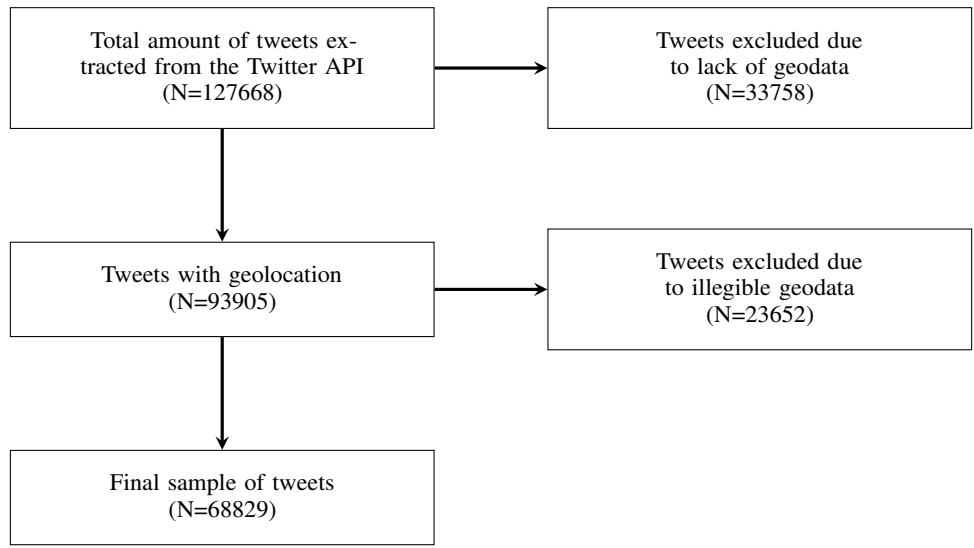
\begin{figure}[h!]
    \centering
    \begin{tikzpicture}[
        node distance=1.5cm,
        start chain=1 going below,
        every join/.style=arrow,
        ]
        \coordinate[on chain=1] (tc);
        \node[mynode, on chain=1] (n2)
        {Total amount of tweets extracted from the Twitter API \\ (N=127668)};
        \node[mynode, join, on chain=1] (n3)
        {Tweets with geolocation \\ (N=93905)};
        \node[mynode, join, on chain=1] (n4)
        {Final sample of tweets \\ (N=68829)};

        \begin{scope}[start chain=going right]
        \chainin (n2);
        \node[mynode, join, on chain]
            {Tweets excluded due to lack of geodata \\ (N=33758)};
            \chainin (n3);
            \node[mynode, join, on chain]
                {Tweets excluded due to illegible geodata \\ (N=23652)};
        \end{scope}
    \end{tikzpicture}
    \caption{Flowchart of data collection/filtration}
    \label{fig:flowchart_tweets}
\end{figure}

The process of filtering and excluding irrelevant tweets from the initial data collection is illustrated as a flowchart in figure \ref{fig:flowchart_tweets}. The total amount of tweets initially collected was 127668, based on the search query described earlier in this section. However, out of these tweets, 33758 did not contain any location in the tweet field nor in the user field, meaning that they were excluded from the data set. Next, a number of tweets did not contain a legible location (e.g. 'the couch' provides no actual spatial information of where a user is based), further excluding 23652 tweets from the data set. The final data set thus contained roughly half of the initial data set (68829 tweets).

A special feature which Twitter provides its users is that of a 'retweet'. A retweet is the re-post of a tweet by a different user on Twitter, and the sentiment expressed through such an act is not obvious. In the context of this article, we consider a retweet as a simple echoing of the original tweet in which the original author is credited (likes of a retweet counts towards the original). Since there is no text or information added by the user retweeting, we assumed that this meant that the user retweeting is in agreement with the original statement. However, an issue arose when retweets exceed 140 characters. Twitter has changed the maximum length of tweets multiple times, however to assure backward compatibility a retweet is truncated at the original 140, and this is the way they appear once fetched by the Twitter API. This was foreseen to potentially cause problems with our model, as the sentiment might be hard to gauge, especially since most of the 140 characters always include a 'RT :', as well as the original username and often multiple URLs. To handle this issue, we isolated all so called 'unfulfilled' retweets, that is tweets specifically starting with 'RT : @' and then a username which is limited to 15 characters, and which ends with '...' or the specific character '…' \footnote{("U+2026" in Unicode)}. We then searched through the whole data set, to see if the original tweet, which has been retweeted is in the data set. If this is the case the truncation is replaced with the remainder of the tweet. The user retweeting is still accredited as author, and nothing else is changed. If the original tweet is not found in the data set (perhaps this user does not have an illegible location tied to the account), then the tweet is dropped from the data set. Twitter also has the feature of a 'quote tweet' in which additional text is added to the original tweet. This we did not count as a retweet but instead as its own tweet.

For the machine-learning based sentiment analysis, the data needed to be cleaned and put in the appropriate format. This included removing emojis, usernames, URLs, as well as stop words and the keywords in our search query since these appear in all tweets. If a tweet, after cleaning, consisted of less than five characters, it was removed from the data set, as it was most likely only a URL or similar which is removed in the clean. Processing code can be found on GitHub\footnote{\url{https://github.com/andebraa/wind_power_analysis}}

\begin{figure}[h!]
    \centering
    \includegraphics[width=0.7\textwidth]{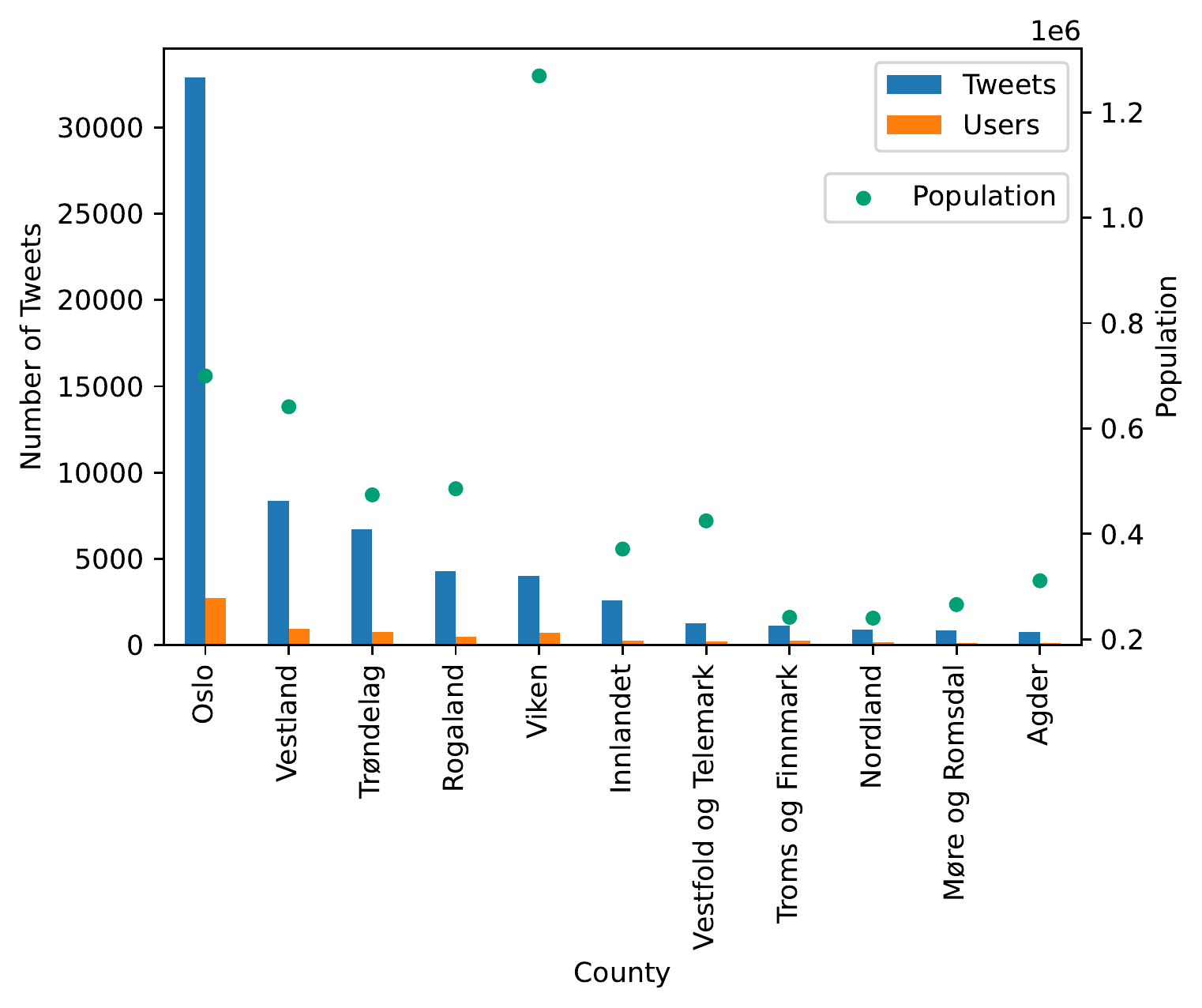}
    \caption{Number of tweets, users and population per county}
    \label{fig:frequency_county}
\end{figure}

The geographical distribution of tweets in our data set was heavily weighted towards the capital region of Oslo, as can be seen in figure \ref{fig:frequency_county}, which holds about 13 percent of the population\footnote{Based on population data from 2022, https://www.ssb.no/statbank/table/07459/}, but 54 percent of all tweets in the data set.

\begin{figure}[h!]
    \centering
    \includegraphics[scale=0.7]{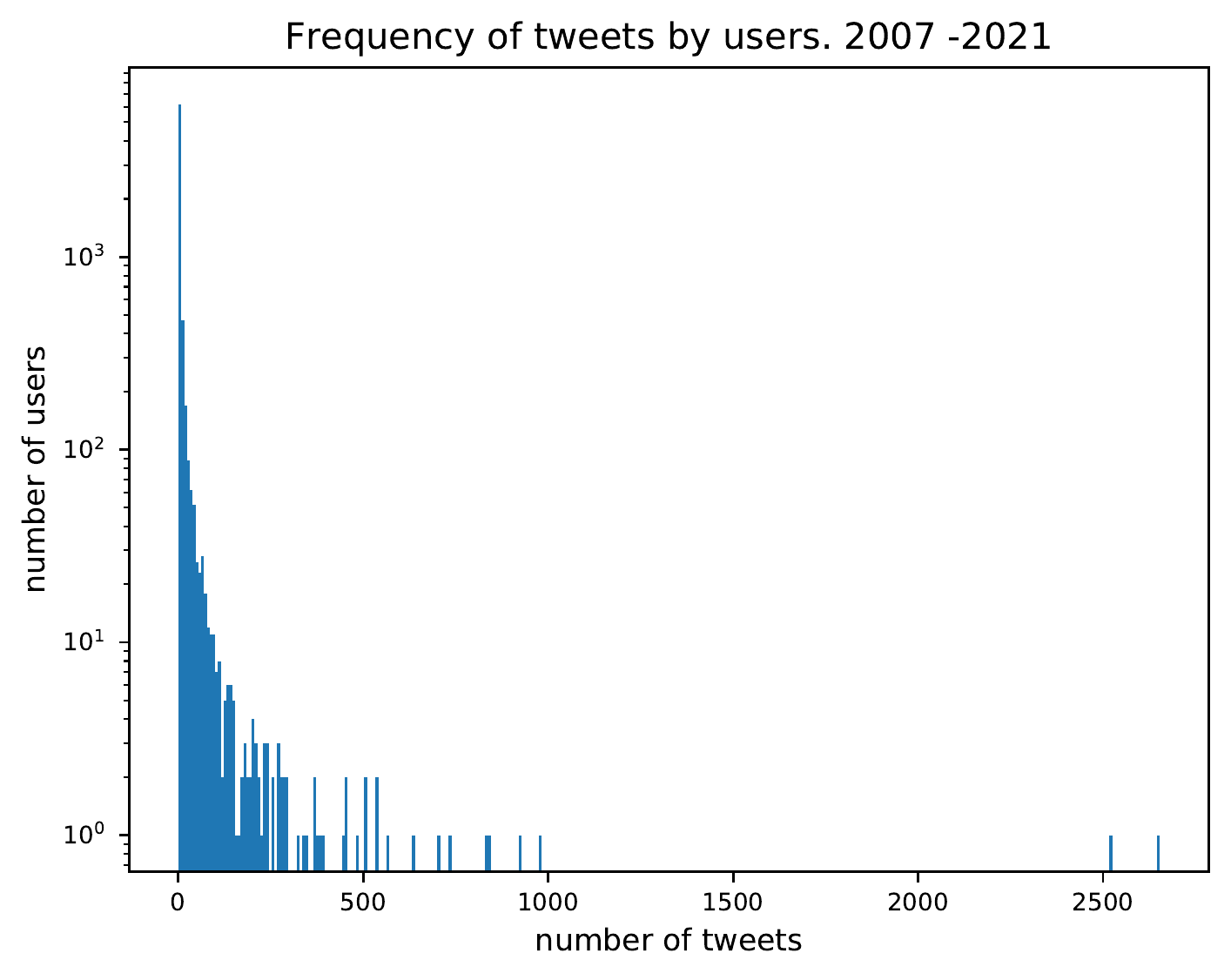}
    \caption{The amount of tweets that occurs in the data set per user}
    \label{fig:frequency_users}
\end{figure}

The data set contained 7287 unique users, with different activity levels on Twitter. Figure \ref{fig:frequency_users} shows the amount of users (on a logarithmic scale) producing a certain amount of tweets. Most users only appear in the data set a few times, but there are a few users with close to 1000, 1500 and 2500 tweets. The most frequent tweeter, with 2647 tweets, is notably the organisation 'La Naturen Leve', founded in 2013 as an anti-wind power organisation\footnote{https://lanaturenleve.no/om-oss/}.

\subsection{Natural Language Processing and NorBERT}
\label{sec:training_model}

Natural Language Processing (NLP) is an automated machine learning approach to allow computers to understand and use human language. A few examples of popular applications of this is voice recognition on mobile devices, translation, and prediction for use in texting apps. Our use of this technology is that of sentiment analysis which extracts subjective quantities from the text, in our case the opinion towards wind power.

The NorBERT model \citep{kutuzovLargeScaleContextualisedLanguage2021} used in this article is a BERT (Bidirectional Encoder Representations from Transformers) model. BERT models are non-causal models that do not generate text but create representations of text and achieve state-of-the-art performance on many NLP tasks. NorBERT was created to work specifically for the Norwegian language and its dialects. It is a joint model meaning it is made to handle both of the official written languages in Norway (Bokmål and Nynorsk). This is an important factor as we wish to include as much of the rural tweets as possible which might be written in the local language variation.

The NorBERT model is trained on five different corpora, including amongst others the C4 web-crawled corpus and Norwegian Newspaper corpus. Despite the pre-training, we still needed our model to pick up on Twitter specific nuances which might not occur as often in the corpora, especially the more official corpora such as the Norwegian newspaper corpus and Wikipedia. This fine-tuning is done manually by iterating through random tweets and assigning it a label 0, 1 or 2 for negative, neutral or positive respectively. As mentioned earlier we chose to do this completely randomly, and on the larger data set which is not limited to only geolocated tweets. We also explore some alternative annotating, by selecting the tweets with the least certain results after the first fine tuning and application of NorBERT. \footnote{This was thought to simply strengthen the model on data which it was unsure about, however it seemed to have little effect on the accuracy of the model.}

The main evaluation metric to evaluate the various machine learning models is that of the F1 score. The F1 score can be used in classification problems and especially on datasets which are skewed in favour of one class \citep{murphyProbabilisticMachineLearning2022}. This is due to the fact that it also considers wrong classifications, and not simply the amount of correct classifications. The F1 score is defined as the harmonic mean of the precision and recall, i.e.

\begin{equation*}
     \textrm{F1} = 2\cdot \frac{\textrm{Precision} \cdot \textrm{Recall}}{\textrm{Precision + Recall}}.
\end{equation*}

Precision is the fraction of true positives to the sum of true and false positives identified by the model while recall is the fraction of true positives to the sum of true positives and false negatives. Since the harmonic mean is more conservative than the arithmetic mean, a high F1 score (close to one) requires both a high precision and high recall.

During fine tuning we found that the ternary data set would perform very poorly, with the positive label having an F1 score of 0.2. We resolved this by combining neutral with either positive or negative tweets. Since most of the training data was neutral, followed by negative, the combination of negative and neutral caused an issue with skewed data sets. This led to the method predicting negative and neutral with a high accuracy since this was the majority of the dataset, however due to the model struggling with positive labels the total F1 score was unusable (close to 0.3),  The average F1 score using three labels negative,
neutral and positive was 0.5, but varying between the different label classes with negative being the easiest to discern at 0.7. The final dataset combines the positive and neutral category, resulting in a negative and a non-negative category, and an F1 score of 0.88.

\begin{table}[h!]
    \centering
    \caption{The number of annotated tweets in the training data set. Training data is pulled from the original data set before geopositioning is performed.}
    \begin{tabular}{cc}
    \toprule
    Category & Number of tweets \\
    \midrule
    negative &  1784\\
    neutral  &  3149\\
    positive &  967 \\
    \bottomrule
    \end{tabular}

    \label{tab:trainingdata_split}
\end{table}

\section{Results}
\label{sec:results}

In this section, we begin with looking at the data without any sentiments analysis applied to it, to identify some general characteristics and temporal patterns. Then, we continue with presenting the results of the machine learning and sentiment analysis. Lastly, we analyse the spatio-temporal development of the data, both quantitatively and qualitatively.

\begin{table}[h!]
\centering
\caption{Overview of Twitter data}\label{table:data_overview}
\begin{tabularx}{\textwidth}{@{\extracolsep{\fill}}lccccc@{\extracolsep{\fill}}}
\toprule
Year & \# of tweets  & \# of new users in data set & \# of active users & share of users being new & tweets per user \\
\midrule
2008 & 16     & 9    & 9    & 1     & 1.78 \\
2009 & 411    & 213   & 218  & 0.98 & 1.89 \\
2010 & 682    & 285  & 349  & 0.82  & 1.95 \\
2011 & 931    & 328  & 455  & 0.72  & 2.05 \\
2012 & 2163   & 654  & 872  & 0.75  & 2.48 \\
2013 & 1910   & 538  & 852  & 0.63  & 2.24 \\
2014 & 2129   & 449  & 800  & 0.56  & 2.66 \\
2015 & 2497   & 431  & 850  & 0.51  & 2.94 \\
2016 & 2560   & 299  & 716  & 0.42  & 3.58 \\
2017 & 2542   & 318  & 740  & 0.43  & 3.44 \\
2018 & 4442   & 413  & 895  & 0.46  & 4.96 \\
2019 & 15783  & 1214 & 2250 & 0.54  & 7.01 \\
2020 & 13288  & 792  & 1985 & 0.40  & 6.69 \\
2021 & 9512   & 714  & 1944 & 0.37  & 4.89 \\
2022 & 9961   & 630  & 1868 & 0.34  & 5.33 \\
\bottomrule \\
\end{tabularx}
\end{table}

Table \ref{table:data_overview} provides a general overview of the Twitter data and the development over time, before any sentiment classification is introduced. From this illustration of the data we can see how the activity slowly builds up before it takes off in 2019, both in terms of number of tweets and active users. Additionally, we see that there is a diminishing share of users tweeting for the first time (related to wind power) as time goes on, indicating that the pool of people posting tweets related to wind power grows larger with time. 2019 is a special year, with numerous new users appearing in the data set as well as high activity in terms of tweets being posted, and the average tweets posted per user. The activity remains at the same high level in 2020, although there are fewer new users entering the conversation. For 2021 and 2022, the activity decreases even if the number of active users remain more or less the same, however it remains high compared to the pre-2019 level\footnote{It should be noted that the final data collection occurred in October 2022, meaning that there is some data from 2022 that escape our analysis.}.

\subsection{Temporal development of wind power opposition}
Public opinion is dynamic and may change over time with different developments in society. Many factors have been identified to influence public sentiment towards wind energy, such as perceived ecological and visual impacts as well as ownership structure. Additionally, there are country-specific differences that may influence the public opinion \citep{vuichardKeepItLocal2022}. As mentioned, we see significant increase in activity over time, especially from 2019 and onwards, when also the amount of installed wind power increased.

\begin{figure}[h!]
    \centering
    \includegraphics[width=\textwidth]{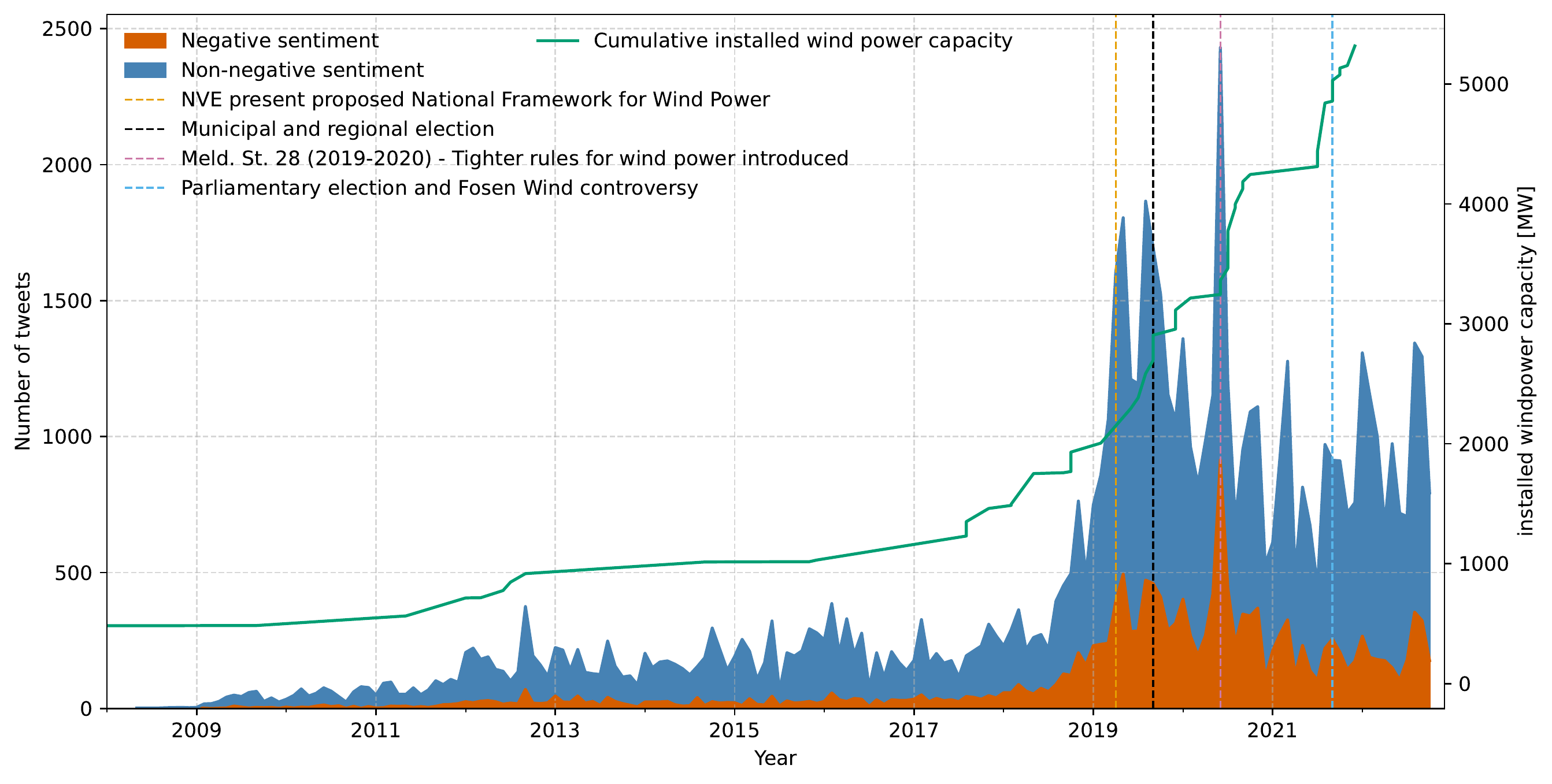}
    \caption{Temporal development of sentiment expressed on Twitter (monthly aggregation), installed wind power capacity and possible influential events}
    \label{fig:temporal_development_general}
\end{figure}

In figure \ref{fig:temporal_development_general}, we see the temporal development of sentiment expressed on Twitter, starting with generally low amounts of tweets up until November 2018, where we for the first time have more than 500 tweets in one month. From 2019 and onwards, both the twitter activity and the installed wind power capacity increase. The first large spike in activity occurs in April/May 2019, where the activity is more than threefold compared to December 2018. This peak corresponds to the time when NVE presented their proposed NFWP (1 April 2019) \citep{nveForslagTilNasjonal2019}. Similarly, we see a spike in August 2019, a month before the municipal and regional election as well as in June 2020, during which the Government announced that the licencing process for wind power would be stricter and controversies around wind power development on the island Haramsøya, Møre og Romsdal received attention. After June 2020, the activity stays between 500–1400 tweets per month.

\begin{figure}[h!]
    \centering
    \includegraphics[width=0.8\textwidth]{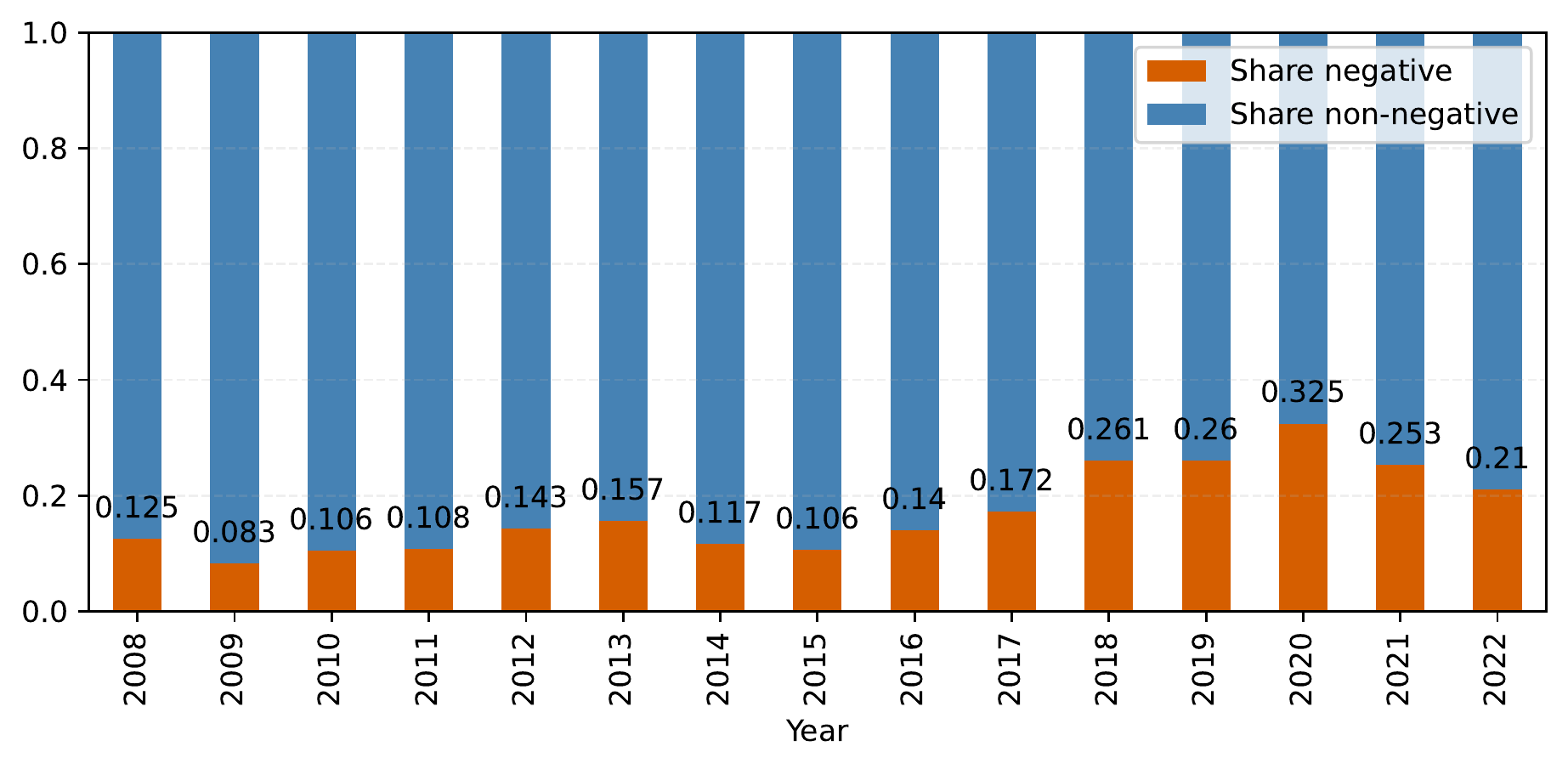}
    \caption{Share of tweets by category for 2008-2022, on a yearly basis}
    \label{fig:share_negative_time}
\end{figure}

\begin{figure}[h!]
    \centering
    \includegraphics[width=0.8\textwidth]{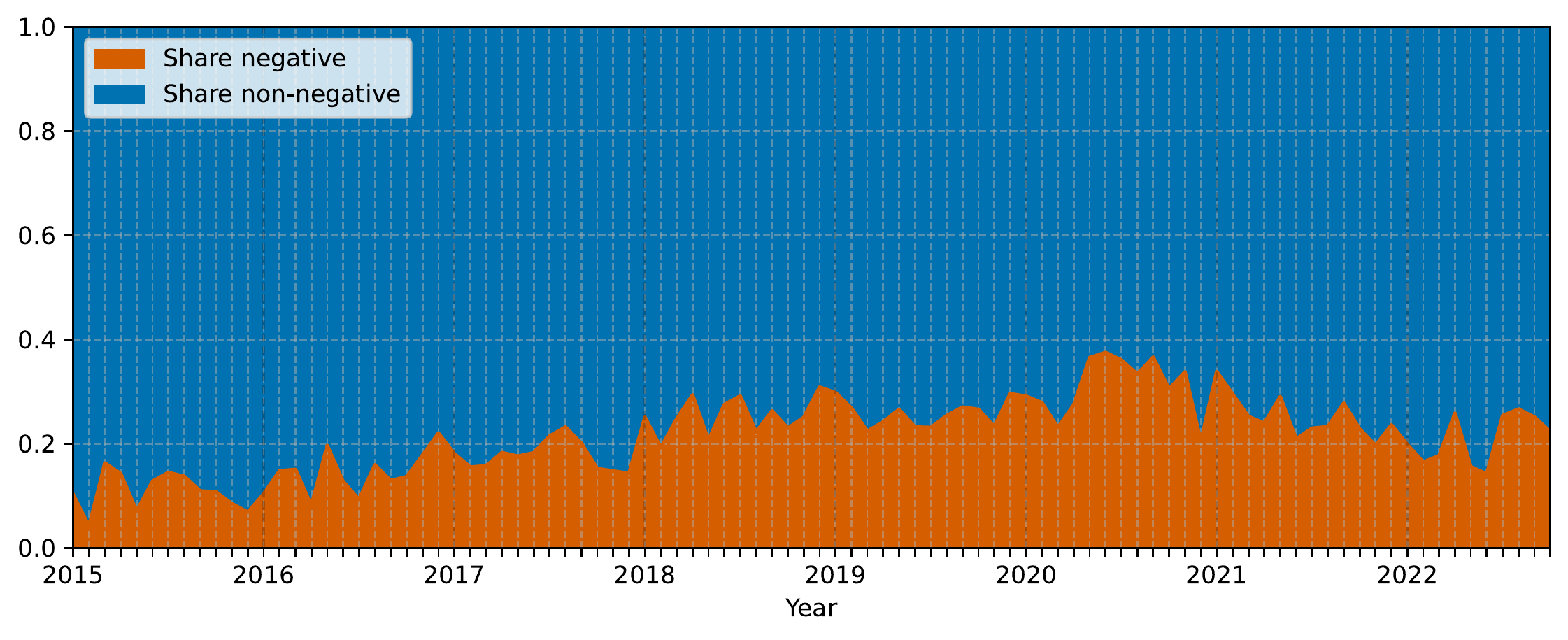}
    \caption{Share of tweets by category for 2015-2022, on a monthly basis}
    \label{fig:share_negative_month}
\end{figure}

From studying the data on an aggregated annual basis, we can see in figure \ref{fig:share_negative_time} how the share of negative tweets begin to increase beyond previous small fluctuations in 2017, reaching its peak in 2020 with 32.5\% of tweets categorised as negative. After 2020, the share of negative sentiment appears to be on a decreasing trend. We are furthermore able to explore the temporal changes in more detail, at a monthly resolution. Figure \ref{fig:share_negative_month} shows the share of tweets by category between 2015 and 2022. The share of negative sentiment does not fluctuate very drastically, and seems to follow the yearly values, with the highest share of negative sentiment in June 2020. After this peak the share of negative tweets decrease somewhat and by 2022 show fluctuations at a generally lower level.

\begin{figure}
    \centering
    \includegraphics[width=0.8\textwidth]{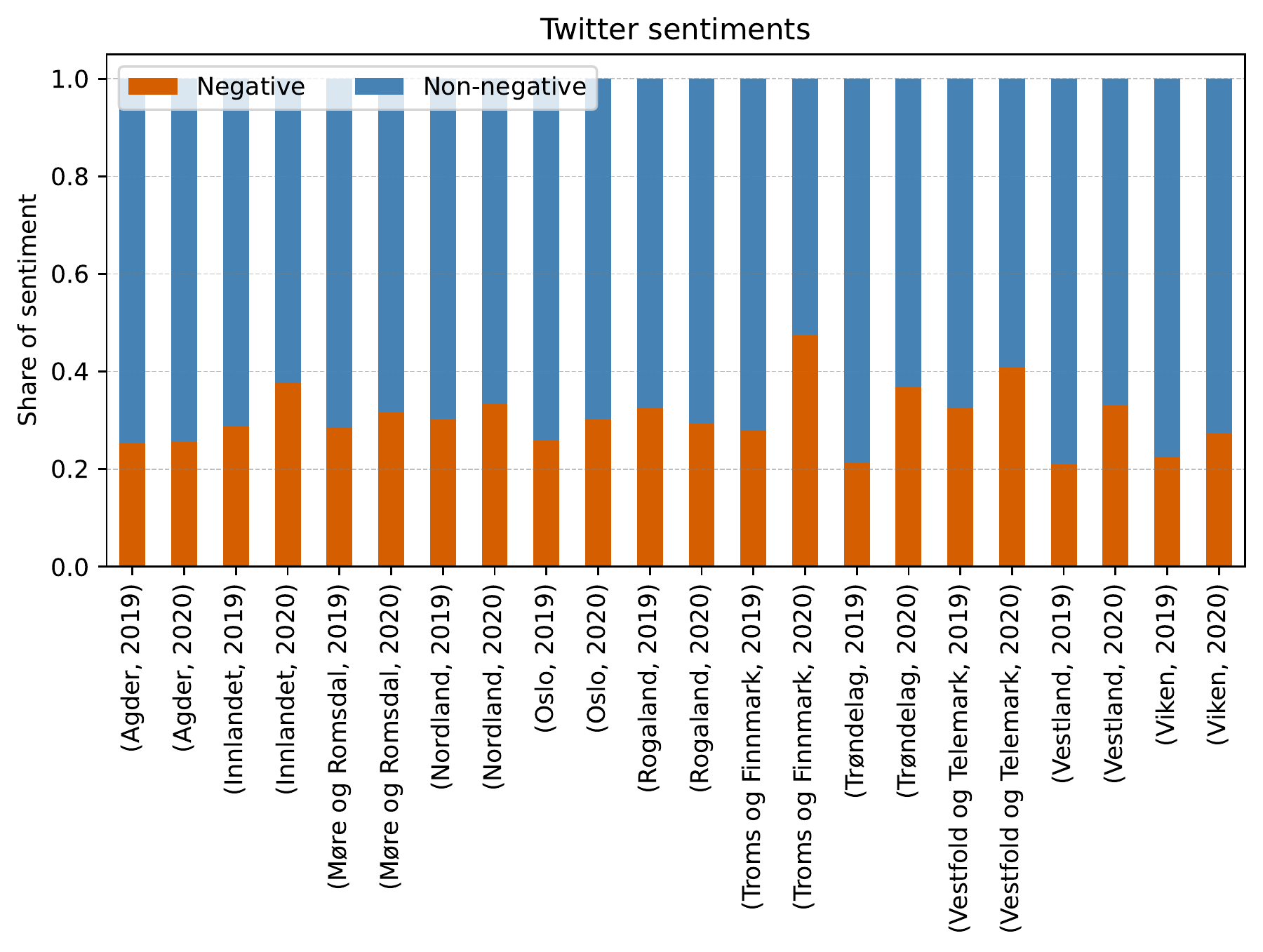}
    \caption{Regional share of tweet by category for 2019 and 2020}
    \label{fig:spatio-temp-1920}
\end{figure}

Up until now, we have analysed the results on an aggregated country level, but the results are also available at a higher spatial resolution. In figure \ref{fig:spatio-temp-1920}, we see the share of tweets by category for both 2019 and 2020 for each region (NUTS3) in Norway. From the results we can see regional differences in the share of negative tweets as well as diversions in terms of change between 2019 and 2020. In some regions the year-on-year change is only a few percentage points, while in for example Vestland and Trøndelag the change is 12 and 15 percentage points, respectively. Visual representations of the regional differences can be found in appendix \ref{app:heat_maps} and figure \ref{fig:county_sentiment}.

\subsection{Comparative spatio-temporal results}
In addition to only assessing our results, we compare the annual results and trends with survey research by CICERO \citep{aasenFolkOgKlima2022}\footnote{To extract data from the graphs presented by \citet{aasenFolkOgKlima2022}, we used the tool WebPlotDigitizer Version 4.6 \citep{rohatgiWebplotdigitizerVersion2022}, and the data may as such be slightly inaccurate.} and University of Bergen \citep{ivarsflatenNorwegianCitizenPanel2020,ivarsflatenNorwegianCitizenPanel2022}. The surveys of \citet{ivarsflatenNorwegianCitizenPanel2020,ivarsflatenNorwegianCitizenPanel2022} are specific rounds within the \textit{Norwegian Citizen Panel}, which is an internet-based survey of Norwegian citizens' attitudes in societal and political questions. The panel is randomly recruited from the National Population Register and included 12904 responses in wave 16 \citep{hogestolNorwegianCitizenPanel2019}, and 13697 in wave 22 \citep{skjervheimNorwegianCitizenPanel2021}. The number of respondents who were asked whether they agreed that Norway should build more onshore wind power plants was 1728 and 1965 for wave 16 and 22 respectively\footnote{We have chosen to aggregate different strong responses (e.g. agree strongly, agree somewhat and agree) into one category for easier comparison with our data}. Similarly, the results of \citet{aasenFolkOgKlima2022} are based on internet-based surveys within the project \textit{ACT - from targets to action}, organised by the research institute CICERO, with data collected for the years 2018-2021. The sample was slightly above 4000 respondents for each round of data collection, with between 2006 and 2477 replies to whether Norway should increase its onshore wind power generation.

\begin{figure}[h!]
    \centering
    \includegraphics[width=0.8\textwidth]{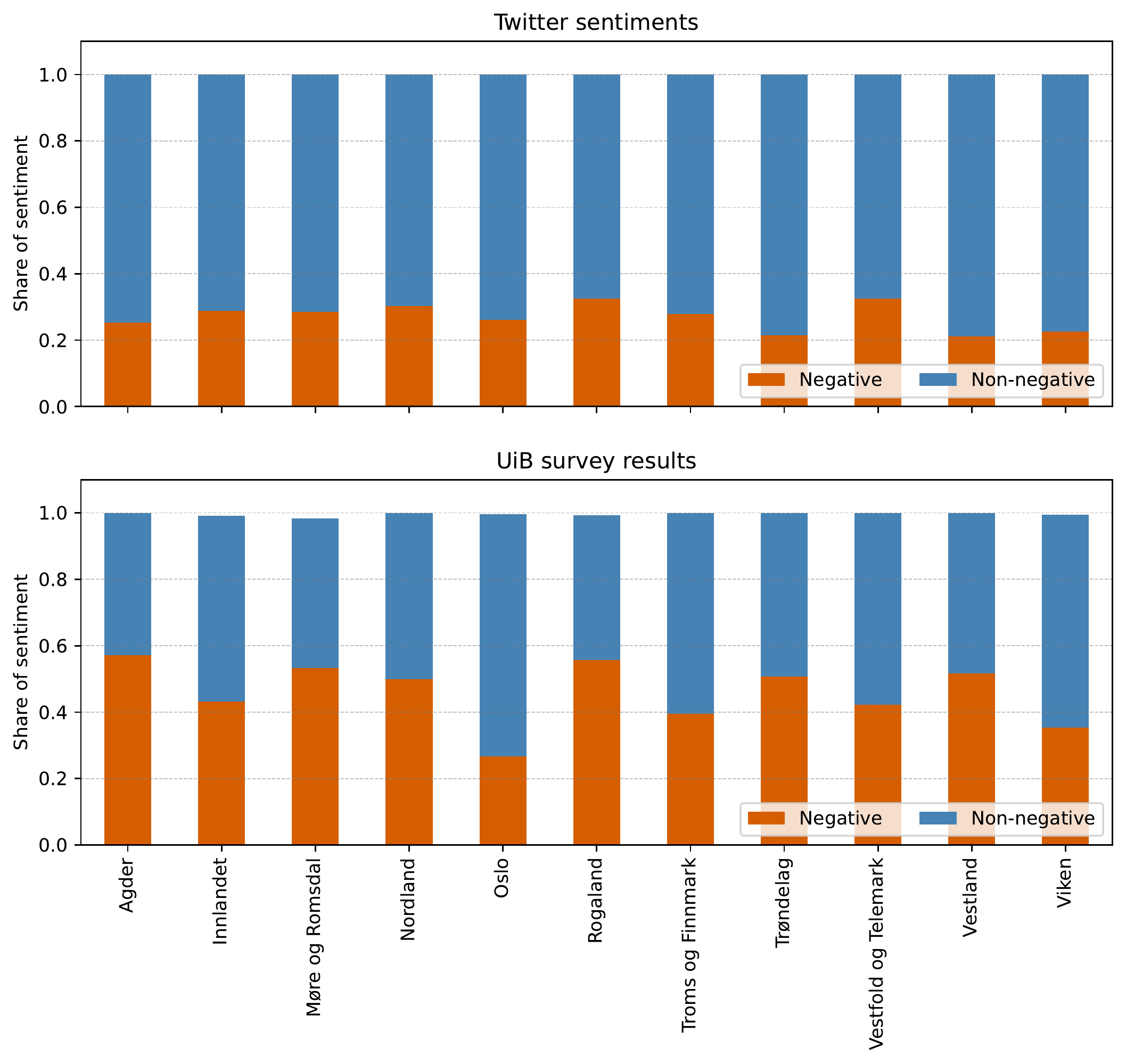}
    \caption{Spatial variation in support for wind power in 2019, on Twitter and according to surveys \citep{ivarsflatenNorwegianCitizenPanel2020}}
    \label{fig:comparison_uib_twitter}
\end{figure}

From figure \ref{fig:comparison_uib_twitter}, we see the difference in the share of negative/non-negative tweets for the different regions of Norway in 2019, both for the results from \citet{ivarsflatenNorwegianCitizenPanel2020} and our results from Twitter. We can see an overall lower share of negative sentiment on Twitter, for all regions other than Oslo in 2019. As seen in figure \ref{fig:frequency_county}, the number of tweets from each region is largely different, and very minimal for Agder, Møre og Romsdal, Nordland, Troms og Finnmark and Vestfold og Telemark, with between 107 and 265 tweets in 2019.

\begin{figure}[h!]
    \centering
    \includegraphics[width=0.8\textwidth]{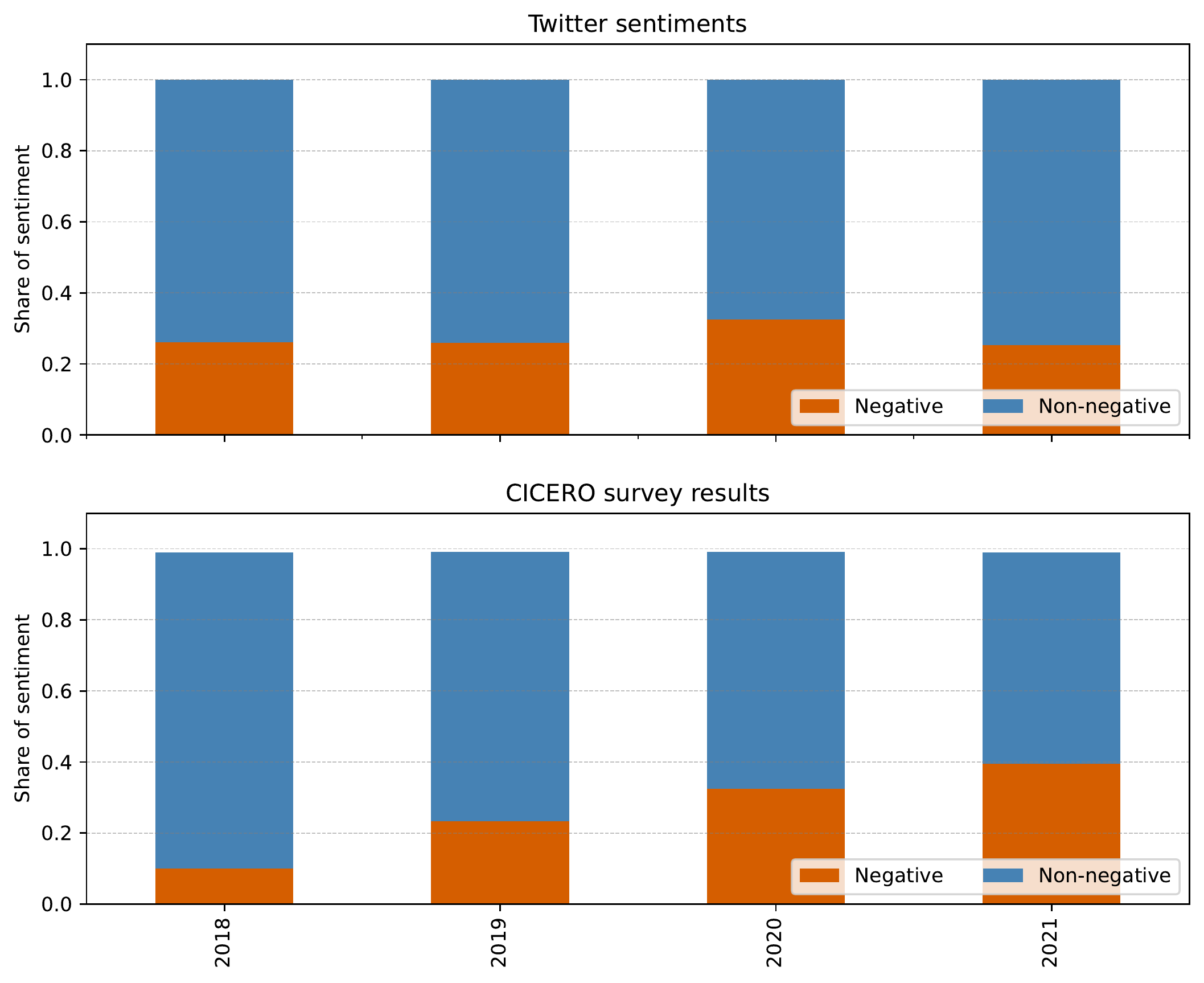}
    \caption{Differences in sentiments expressed on Twitter and in surveys conducted by CICERO for the years 2018-2021 \citep{aasenFolkOgKlima2022}}
    \label{fig:temporal_diff_cicero}
\end{figure}

With the first survey data from 2018, and the latest in 2021, the survey results are much more limited in temporal extent, compared to the Twitter data, limiting our comparison to only those four years. In figure \ref{fig:temporal_diff_cicero}, we see the country-wide annual changes in the share of negative sentiment between 2018-2021 both on Twitter and reported results by \citet{aasenFolkOgKlima2022}. For these four years, the share of negative sentiment on Twitter was already relatively high (almost a 10 percentage points increase between 2017 and 2018), and would have seen a larger difference if the first year of comparison was in 2017. The share of negative sentiment was similar throughout the four years, other than a 6.5 percentage point increase in negative sentiment in 2020. The survey results on the other hand show a much clearer trend of increasing negative sentiment over time, although the year-on-year change is slowed down over time (13.4, 9.1 and 7 percentage point increase in share of negative sentiment).

\begin{figure}[h!]
    \centering
    \includegraphics[width=0.8\textwidth]{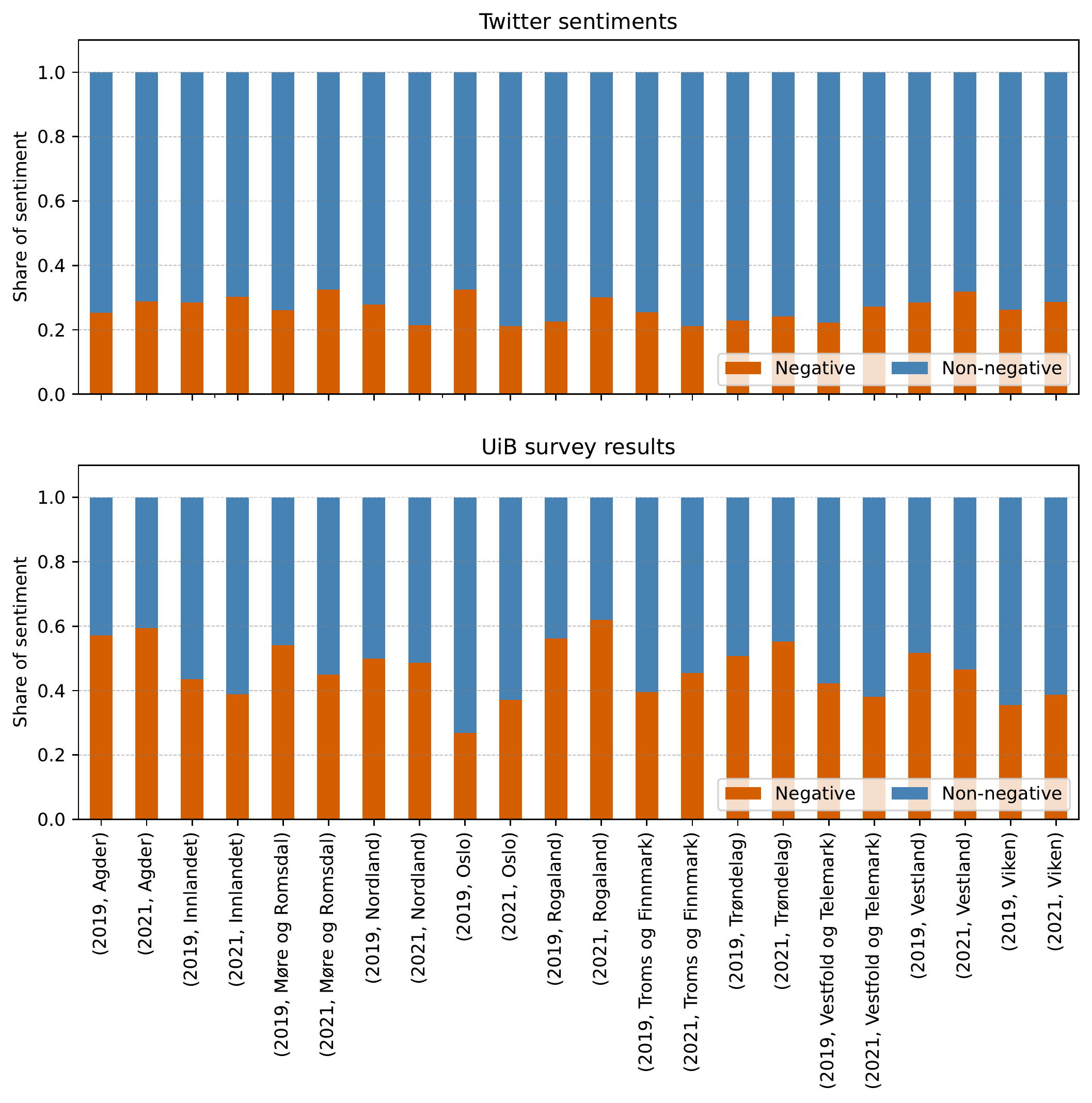}
    \caption{Differences in sentiments expressed on Twitter and in surveys conducted by University of Bergen for the years 2019 and 2021 \citep{ivarsflatenNorwegianCitizenPanel2020,ivarsflatenNorwegianCitizenPanel2022}}
    \label{fig:temporal_diff_uib}
\end{figure}

Similarly, we can compare how sentiments have changed over time in different regions. Figure \ref{fig:temporal_diff_uib} provides two separate data points for each county in Norway, one for 2019 and one for 2021. In addition to generally lower negative sentiment on Twitter, we can see a few reversed trends in the illustration. For example, between 2019 and 2021, the share of negative sentiment increased in Oslo according to survey results, with an opposite change on Twitter. Similarly, the share of negative sentiment decreased in Vestland between 2019 and 2021 in surveys, but increased on Twitter. However, being limited to only two data points for each county only allows for limited analysis. For example, seven out of the 11 counties have their peak in share of negative sentiment on Twitter in 2020, which is not captured by this comparison. Further details on the spatio-temporal development of sentiments for each Norwegian county is outlined in appendix \ref{app:spatio-temporal}.

\section{Discussion}
\label{sec:discussion}

Twitter as a communication platform for wind power discussions has received increasing attention in Norway from 2018/2019 and onwards, when the number of users active in the discussion, and the number of tweets increased more than two- and three times over respectively. This drastic increase likely related to the general increase in wind power developments and particularly controversies and media attention following the release of the NFWP, and one might expect a more significant change in particularly negative sentiments. However, from figure \ref{fig:share_negative_time} and figure \ref{fig:share_negative_month}, we can see that both the yearly and monthly share of negative tweets are not drastically changing after 2018, when the activity and number of new users increases. As such, our data indicates that with the increasing attention to wind power controversies, there was not \textit{only} an influx of opposers to Twitter who changed the discussion and critiqued wind power, but also people who were positive, or at least non-negative. Given the large negative attention and protests against wind power in this time period, one may have expected a more significant increase in negative sentiments and fewer non-negative tweets.

Between 2019 and 2021, there were several important events which one may have expected to drive activity on Twitter, such as the release of the NFWP, the municipal and regional election of 2019 and the suggestion for stricter rules for wind power deployment by the Government in 2020 (Meld. St. 28 (2019–2020)\footnote{\textit{Meldinger til Stortinget (Meld. St.)} are documents that the government sends to the parliament to be discussed, but without proposing a particular decision}). Surprisingly, while there is a large amount of activity at the time when the NFWP was released, it is not overwhelmingly negative. The share of negative tweets in April 2019 is below the yearly average and the whole year of 2019 only fluctuate by a few percentage points. For 2020, the number of tweets and active users are slightly below 2019, but the share of negative tweets continued to increase, hitting an all-time high. June 2020 is the month with the highest number of tweets, negative tweets and share of negative tweets (37.5\%). Exploring the word frequency in this specific month shows mentions of, among other things, the island Haramsøya, situated in Møre og Romsdal, which have attracted significant attention in the public discourse. The Ministry of Petroleum and Energy (OED) approved the plans for 66 MW of wind power on this island in March 2020 \citep{norwegianwaterresourcesandenergydirectorateKonsesjonssakHaramVindkraftverk2021}, after which it has attracted significant media attention and protests\footnote{https://www.nrk.no/mr/vindkraft-pa-haramsoya-1.14700225}. In June 2020 the construction started, which was met by protests and road blockages. This is likely a strong contributor to the activity seen on Twitter.

It is too early to say if the discussion on Twitter have reached its peak in 2020 and if we will continue to see a decline in share of negative tweets, but both at the national level and in most regions, the share of negative sentiment declined in 2021 and 2022. This trend is in contrast to the survey results we have analysed, where the negative sentiment have continued to increase, although at a decreasing pace. Greater awareness of wind power as a technology, and increasing shares of wind in the Norwegian electricity mix may very well keep the activity elevated compared to pre-2019 levels.



The Norwegian Twitter landscape is generally less negative to wind power, compared to the general population, as indicated by our comparison to survey research (see e.g. figure \ref{fig:comparison_uib_twitter}). Despite this, we can still identify both national and regional trends of more negative sentiment from 2019 and onwards, and a change in the discussions taking place. One might attribute this lower negativity to the heavy weight of Oslo-users in the aggregated data set, as surveys indicate that it is the least negative region in Norway towards wind power. However, the spatially disaggregated data in appendix \ref{app:spatio-temporal} also show an increase in negative sentiment for Oslo, both in absolute and relative terms. The spatially disaggregated data is more difficult to analyse, due to the large variations in data volume for the different regions. As outlined in figure \ref{fig:frequency_county}, the differences are substantial. In Agder, in the southern part of Norway, the number of tweets is minimal. In total, there are only 770 tweets throughout the whole time period and in the peak month 31 tweets. This makes the data very volatile and sensitive to statistical noise, limiting its usefulness for analysis. Other regions with limited amounts of data include Møre og Romsdal (N=873), Nordland (N=915), Troms og Finnmark (N=1122) and Vestfold og Telemark (N=1257). For the other regions, which have larger amounts of data, we see indications that even if many conflicts around wind power play out on a local level, the discussions on Twitter are less isolated. For example, the county Viken only has one wind power cluster of 54 MW installed capacity, which began its operations in 2019. Despite very little wind power and local conflicts, there are still distinct peaks in the data for Viken (see figure \ref{fig:app_viken_bar} in appendix \ref{app:spatio-temporal}), which then likely rather relate to developments and discussions at the national level, or in other regions. \citet{brunsNorwegianTwittersphere2018} have previously reported that the Norwegian accounts generally are well connected, and that there is little evidence for the existence of so-called 'echo chambers'. We see further indications of weak geographic clustering in our data for the six regions with more considerable data volumes, where the development and sudden increases (peaks) in activity follow very similar patterns, especially when normalised to the total volume of tweets in a region, as seen in figure \ref{fig:app_relative_regional_trends} in appendix \ref{app:regional_trends}. Through a network analysis (see appendix \ref{app:network_analysis}) of Twitter networks, we quantify this relationship between geographic region and Twitter networks and although there is \textit{some} regional clustering, the results show that it is weak (V = 0.142).

For the comparisons between Twitter and survey results, we are restricted both due to the temporal extent of the surveys, but first and foremost due to the lack of user profile metadata. It is as such not possible to fully understand who is represented in our data set and which societal groups are over/underrepresented. However, survey research suggests that more men than women are registered and use Twitter daily in Norway and that activity is higher in the age groups 18-29, 30-39 and 40-49 \citep{ipsosSOSIALEMEDIERTRACKER2022}.

For the performance of the NorBERT language model, we achieve a slightly lower F1 score (0.88) for the binary classification compared to that of \citet{kimPublicSentimentSolar2021} (0.91), and much worse (0.50 contra 0.80) in the case of a ternary classification. Potential reasons as to why NorBERT performs worse than its English counterpart could be more frequent use of informal language (e.g. dialectal words) in the Norwegian Twitter landscape.

One limiting factor to drawing general conclusions from our analysis is the lack of user profile metadata in our data set, resulting in that we do not know how and to what extent the sentiment expressed should be seen as representative of the general population. The results shown, and the conclusions drawn in this study only provides insight into spatio-temporal variations on Twitter. Even if such insights may be useful for understanding public opinions towards renewable energy technologies, they should be treated with care.

Additionally, combining neutral and positive tweets into the category non-negative leads to fewer nuances and lost information in the data set. Exploring the difference between positive and neutral sentiments would have strengthened the insights and allowed for more detailed comparison with survey results, as those studies contained such information.

\section{Summary and conclusion}
In this work, we have studied how sentiments towards wind power have evolved spatio-temporally using Norwegian twitter data and a BERT deep learning language model, NorBERT. Despite not being representative of the general Norwegian population, our results contribute to the knowledge of public sentiment of Twitter users, and can complement existing survey research. Through the application of machine learning based sentiment analysis, we are able to process 68828 individual expressions of sentiment from 7287 unique Twitter users. While model training only allows us to apply a binary classification of tweets with negative and non-negative sentiments, we achieve an F1 score of 0.88.

Our results show that, similar to in the wider political discourse, wind power became a much more relevant topic on Twitter after 2018/2019, generating on average 4 times  more activity than it had previously. While wind power received much negative attention in Norway around this time  with protests and resistance groups forming leading to the withdrawal of the national framework for wind energy, the increase in negative sentiments on Twitter was more moderate. Still, the share of negative sentiment increased on Twitter and reached a peak in 2020 with 32.5\% negative tweets, after which it has declined. Analysing the results spatially proved difficult, due to the limited number of tweets in certain regions. However, while there are different trends for the various regions, we find a weak relationship between geographical regions and Twitter networks, indicating that discussions are country wide and not dominated by specific regional events or developments. Using both regional and national data provides the opportunity to get an insight into specific time intervals and events. For example, the global peak of negative sentiment in June 2020 show importance of the developments in the contested wind power developments at Haramsøya. Similarly, tweets in January 2021 seem to revolve, to some extent around reports of dead eagles caused by wind power plants. As such, while some nuances are lost with the binary classification and limited user profile metadata, new insights can be found with the high temporal resolution of the data.

This is the first study using Twitter data to analyse sentiments towards wind energy and provides ample opportunities to build upon for future research: 1. While natural language processing of Twitter data has been limited spatially in the Norwegian case, more populous countries like Germany or France may provide better application areas. Larger amounts of data may allow for higher spatial resolution, potentially countering some of the challenges with understanding who is represented in the data set. 2. We compared Twitter data with the location of wind energy projects. In future research spatially explicit Twitter data could be combined with other spatial datasets such as scenicness to give a better understanding of sentiments towards wind energy including drivers. 3.  To further advance the mapping and understanding of public sentiments towards key components of future energy systems, other data sources, such as newspapers or other social media sites, should be explored more in detail.

%

\section*{Acknowledgements}
The authors of this work have received funding from various sources and to various extents. O. Vågerö acknowledges funding from Include – Research centre for socially inclusive energy transitions, funded by the Research Council of Norway, project no 294 687. A. Bråte and A. Wittemann acknowledges funding from UiO:Energy. The funding institutions had no involvement in any part of the study and the responsibility lies with the authors.

We would like to thank Andrei Kutuzov at the Department of Informatics, UiO, for the support in understanding, applying and fine-tuning NorBERT for this project.

\section*{Author contributions}
\textbf{Oskar Vågerö:} Conceptualisation, Validation, Formal analysis, Data Curation, Writing - Original Draft, Visualisation, Project administration; \textbf{Anders Bråte:} Methodology, Software, Validation, Formal analysis, Investigation, Data Curation, Writing - Original Draft, Visualisation; \textbf{Alexandra Wittemann:} Methodology, Software, Formal analysis, Investigation, Data Curation, Writing - Original Draft; \textbf{Jessica Yarin Robinson:} Conceptualisation, Methodology, Formal analysis, Resources, Writing - Review \& Editing, Visualisation; \textbf{Natalia Sirotko-Sibirskaya:} Conceptualisation, Methodology, Software, Formal analysis, Writing - Review \& Editing, Supervision, Project administration; \textbf{Marianne Zeyringer:} Conceptualisation, Methodology, Writing - Review \& Editing, Supervision, Project administration, Funding acquisition

\bibliographystyle{unsrtnat}
\bibliography{main.bib}

\appendix

\section{Spatio-temporal development}
\label{app:spatio-temporal}

\subsection{Agder}
\begin{figure}[h!]
    \centering
    \includegraphics[width=0.8\textwidth]{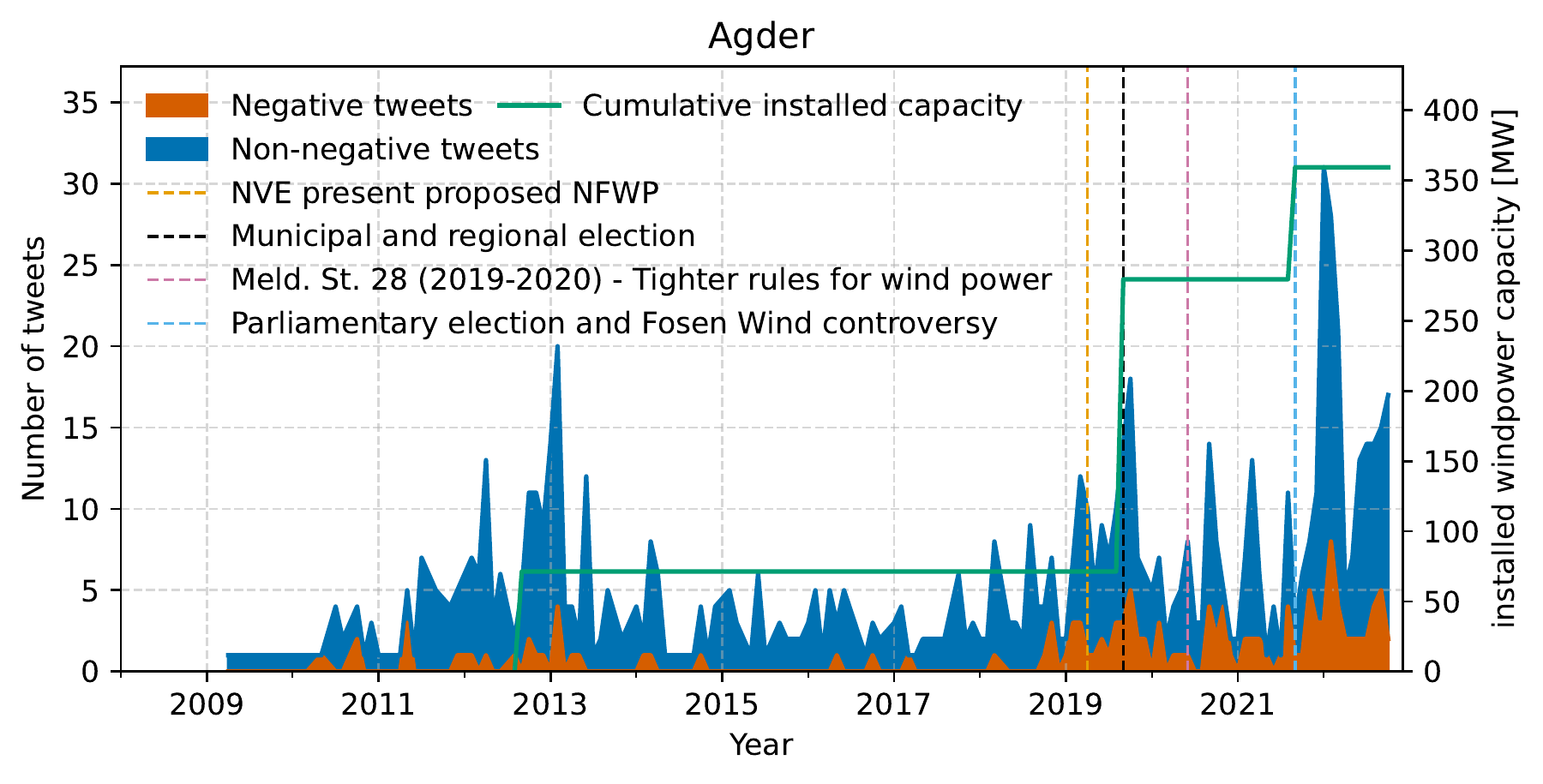}
    \caption{Temporal development of sentiment expressed on Twitter (monthly aggregation), installed wind power
capacity and possible influential events for Agder}
    \label{fig:app_agder_area}
\end{figure}

\begin{figure}[h!]
    \centering
    \includegraphics[width=0.8\textwidth]{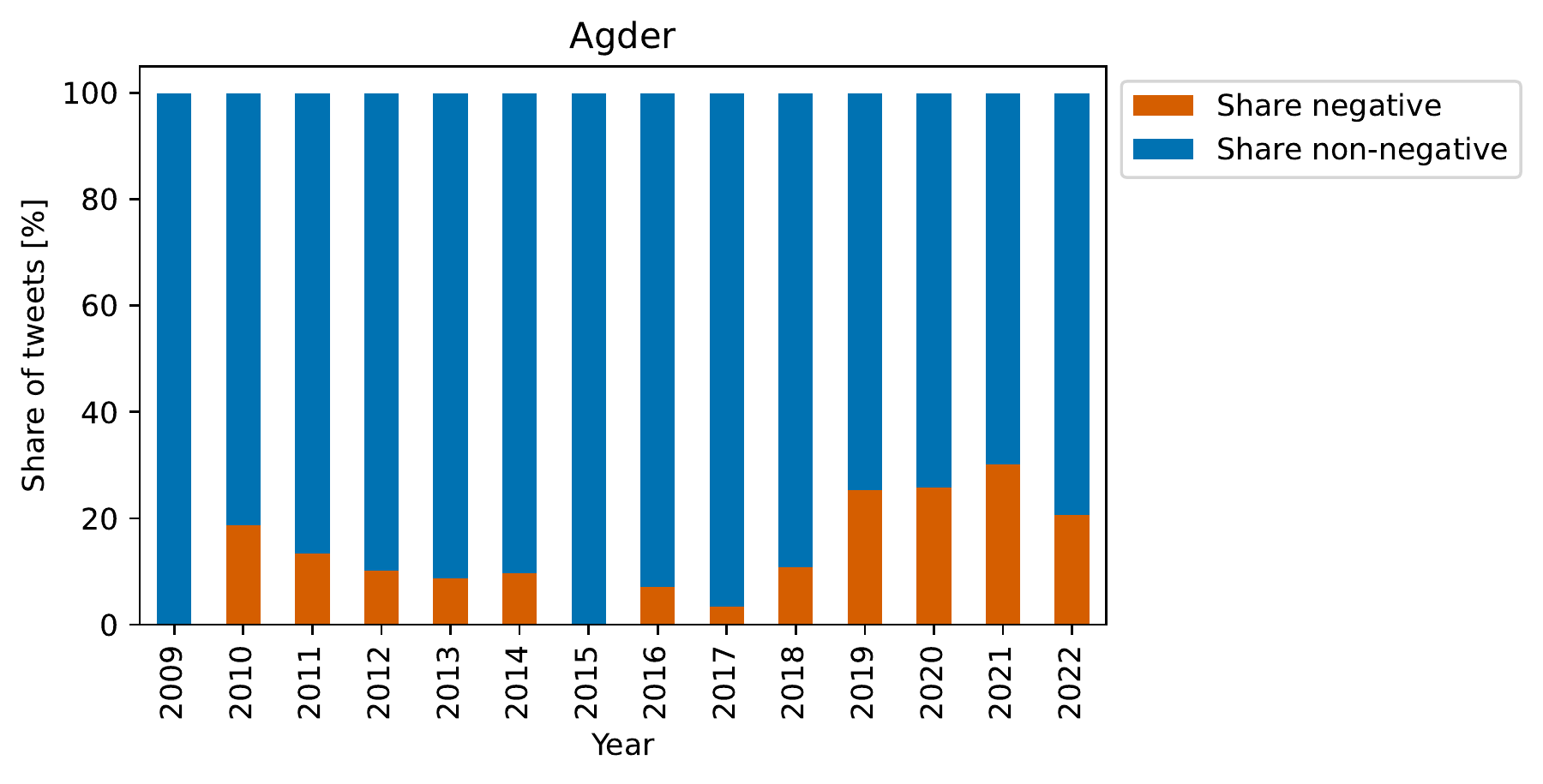}
    \caption{Share of positive and negative tweets per year for Agder (N=770)}
    \label{fig:app_agder_bar}
\end{figure}

\newpage
\subsection{Innlandet}
\begin{figure}[h!]
    \centering
    \includegraphics[width=0.8\textwidth]{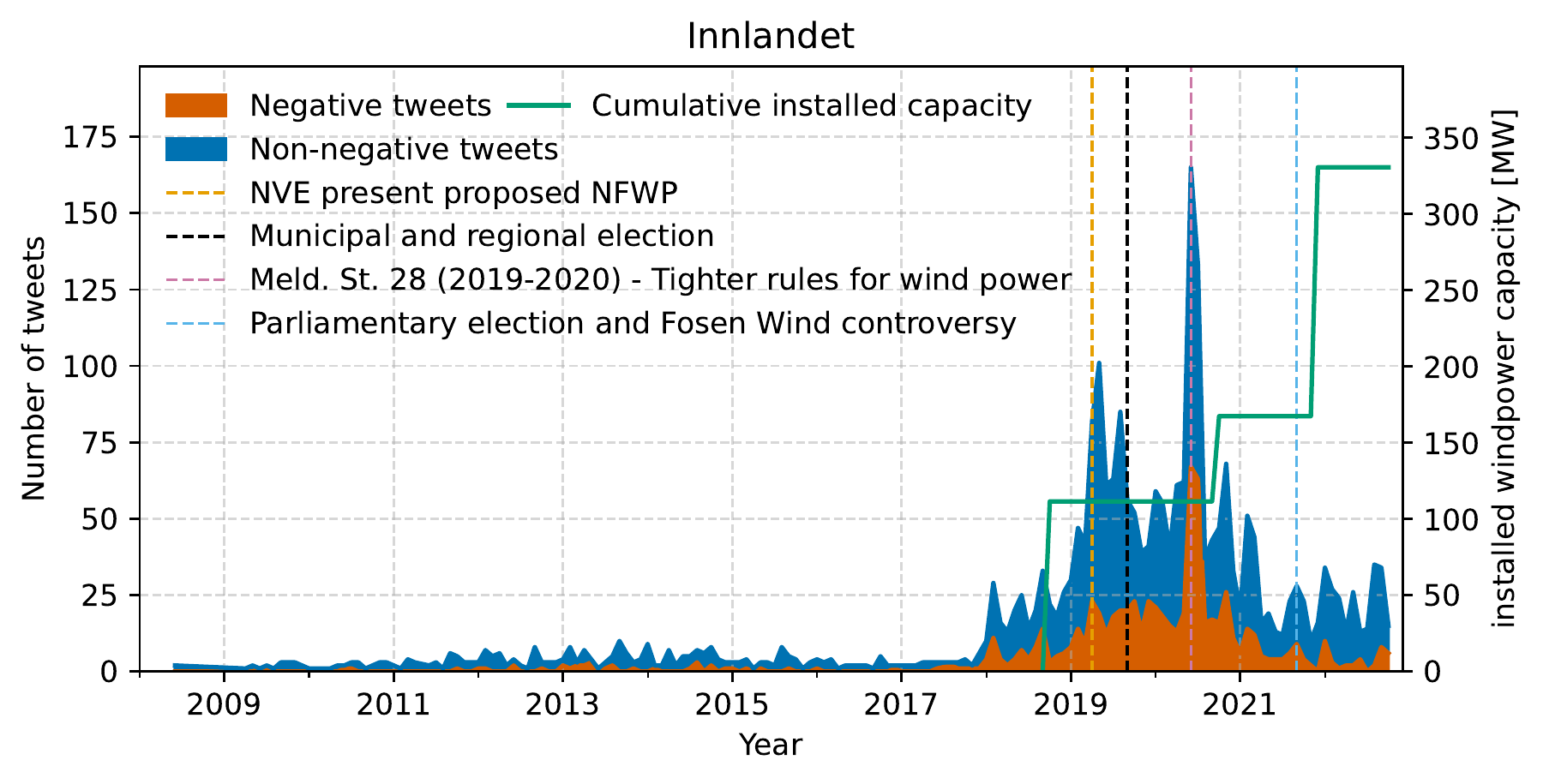}
    \caption{Temporal development of sentiment expressed on Twitter (monthly aggregation), installed wind power
capacity and possible influential events for Innlandet}
    \label{fig:app_innland_area}
\end{figure}

\begin{figure}[h!]
    \centering
    \includegraphics[width=0.8\textwidth]{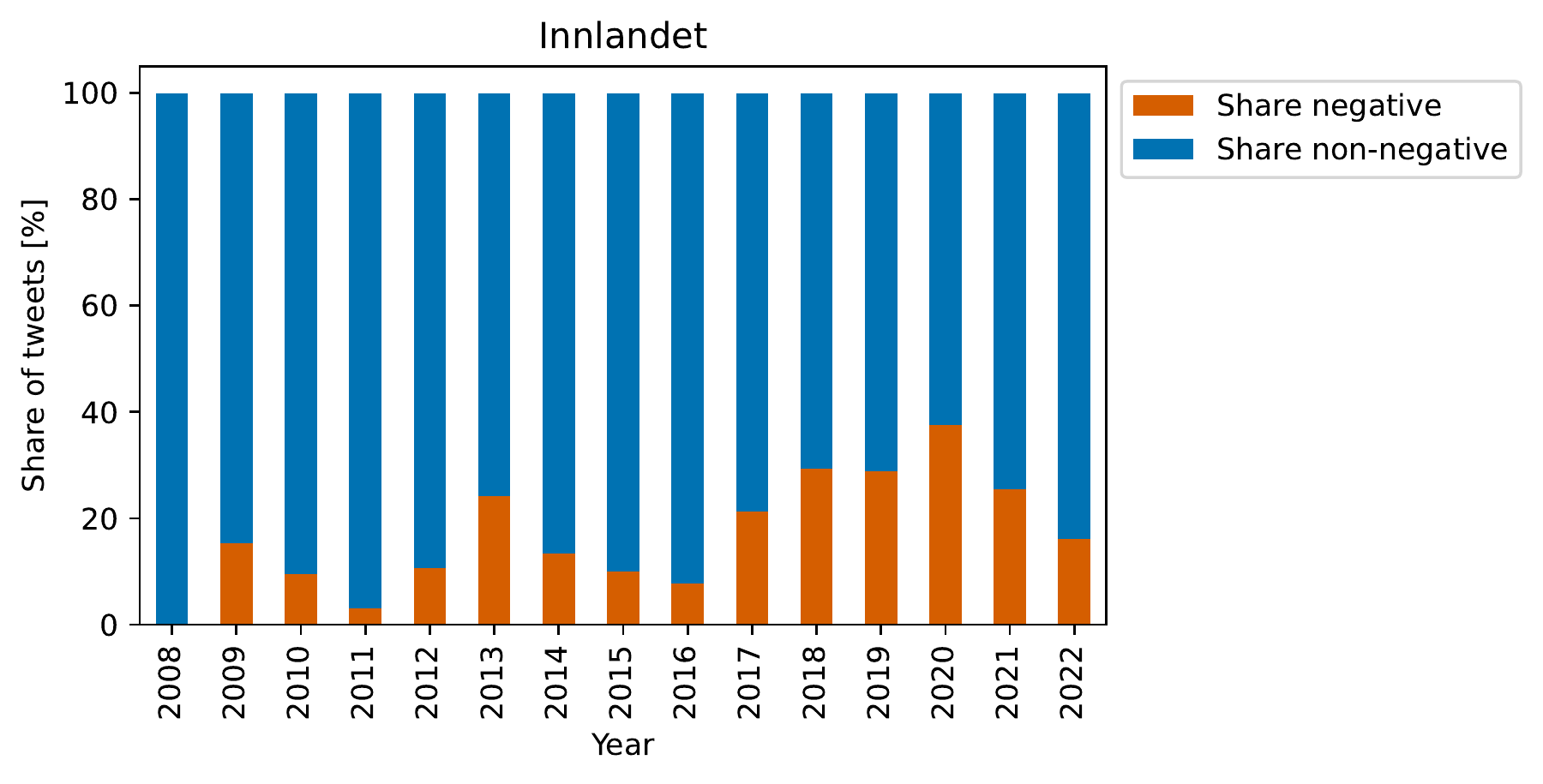}
    \caption{Share of positive and negative tweets per year for Innlandet (N=2591)}
    \label{fig:app_innland_bar}
\end{figure}

\newpage
\subsection{Møre og Romsdal}
\begin{figure}[h!]
    \centering
    \includegraphics[width=0.8\textwidth]{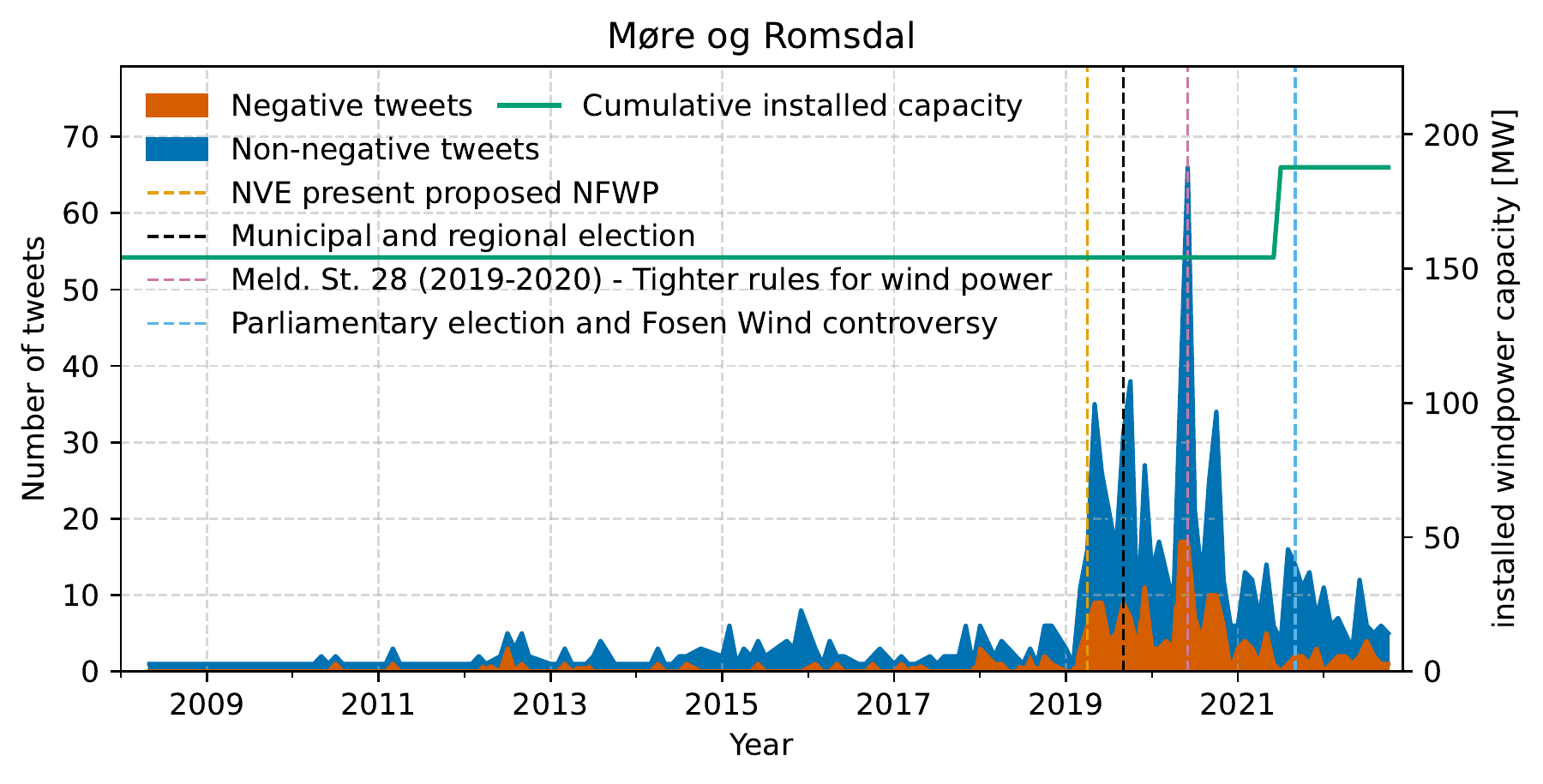}
    \caption{Temporal development of sentiment expressed on Twitter (monthly aggregation), installed wind power
capacity and possible influential events for Møre og Romsdal}
    \label{fig:app_more_area}
\end{figure}

\begin{figure}[h!]
    \centering
    \includegraphics[width=0.8\textwidth]{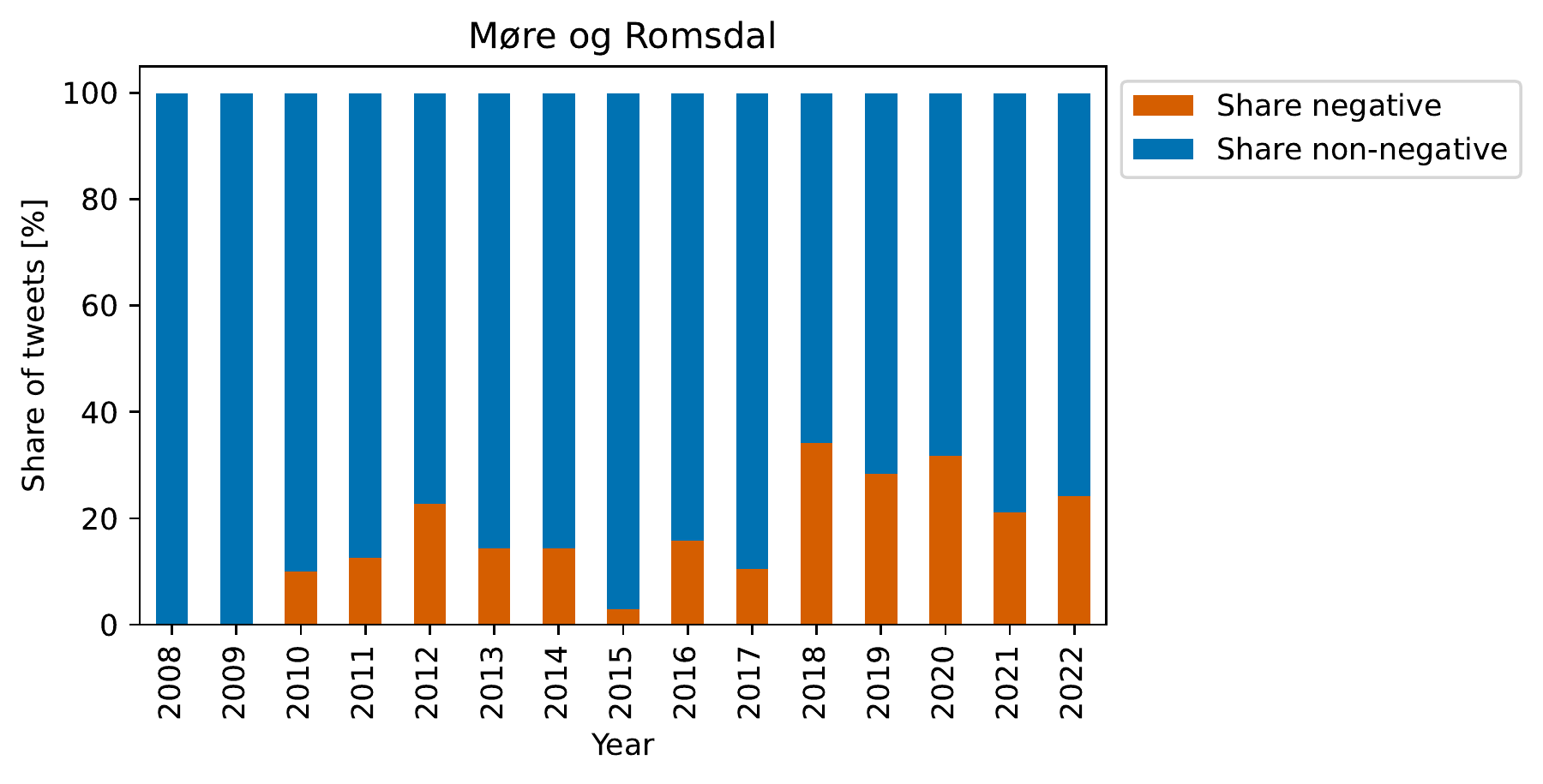}
    \caption{Share of positive and negative tweets per year for Møre og Romsdal (N=873)}
    \label{fig:app_more_bar}
\end{figure}

\newpage
\subsection{Nordland}
\begin{figure}[h!]
    \centering
    \includegraphics[width=0.8\textwidth]{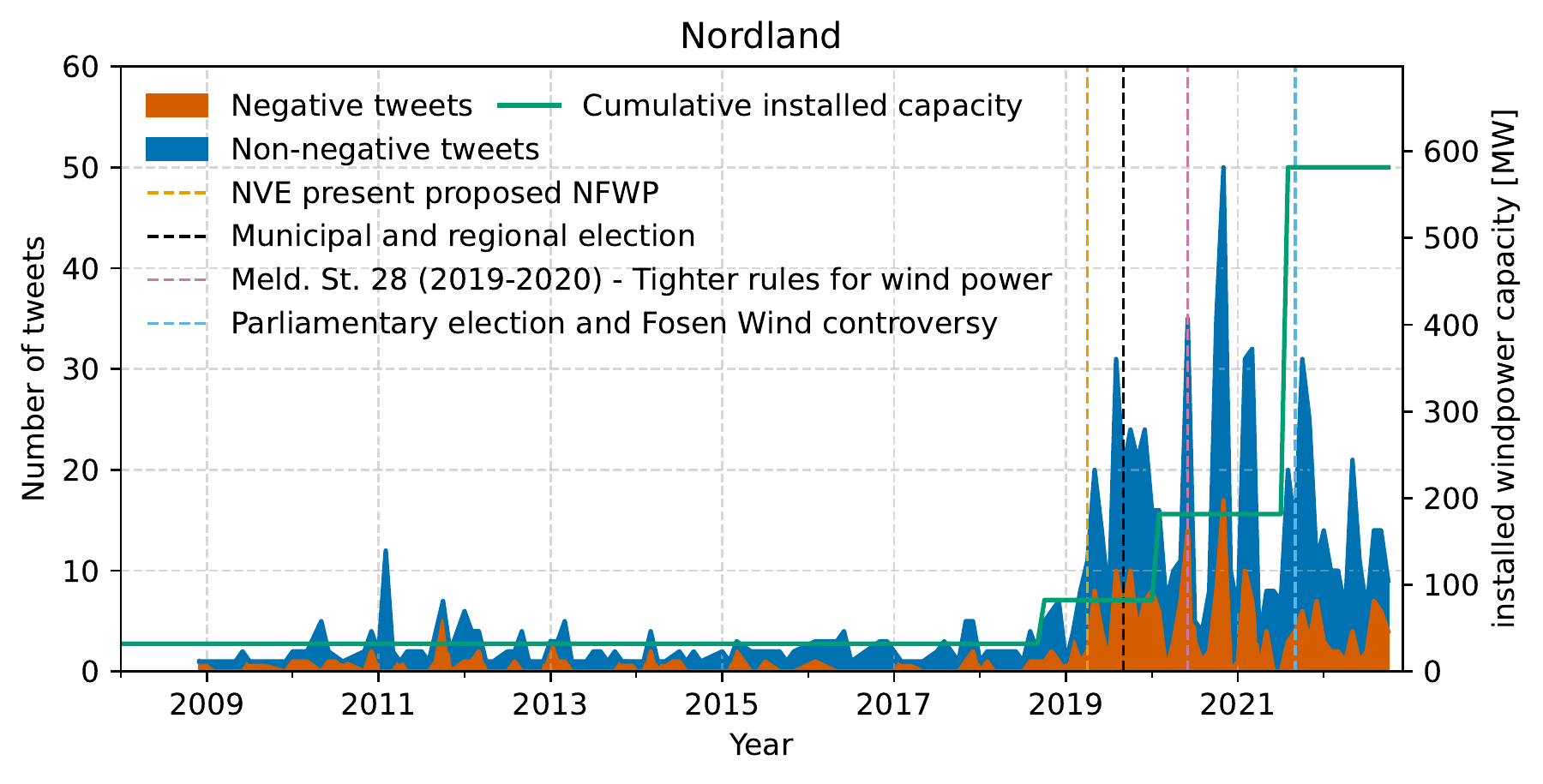}
    \caption{Temporal development of sentiment expressed on Twitter (monthly aggregation), installed wind power
capacity and possible influential events for Nordland}
    \label{fig:app_nord_area}
\end{figure}

\begin{figure}[h!]
    \centering
    \includegraphics[width=0.8\textwidth]{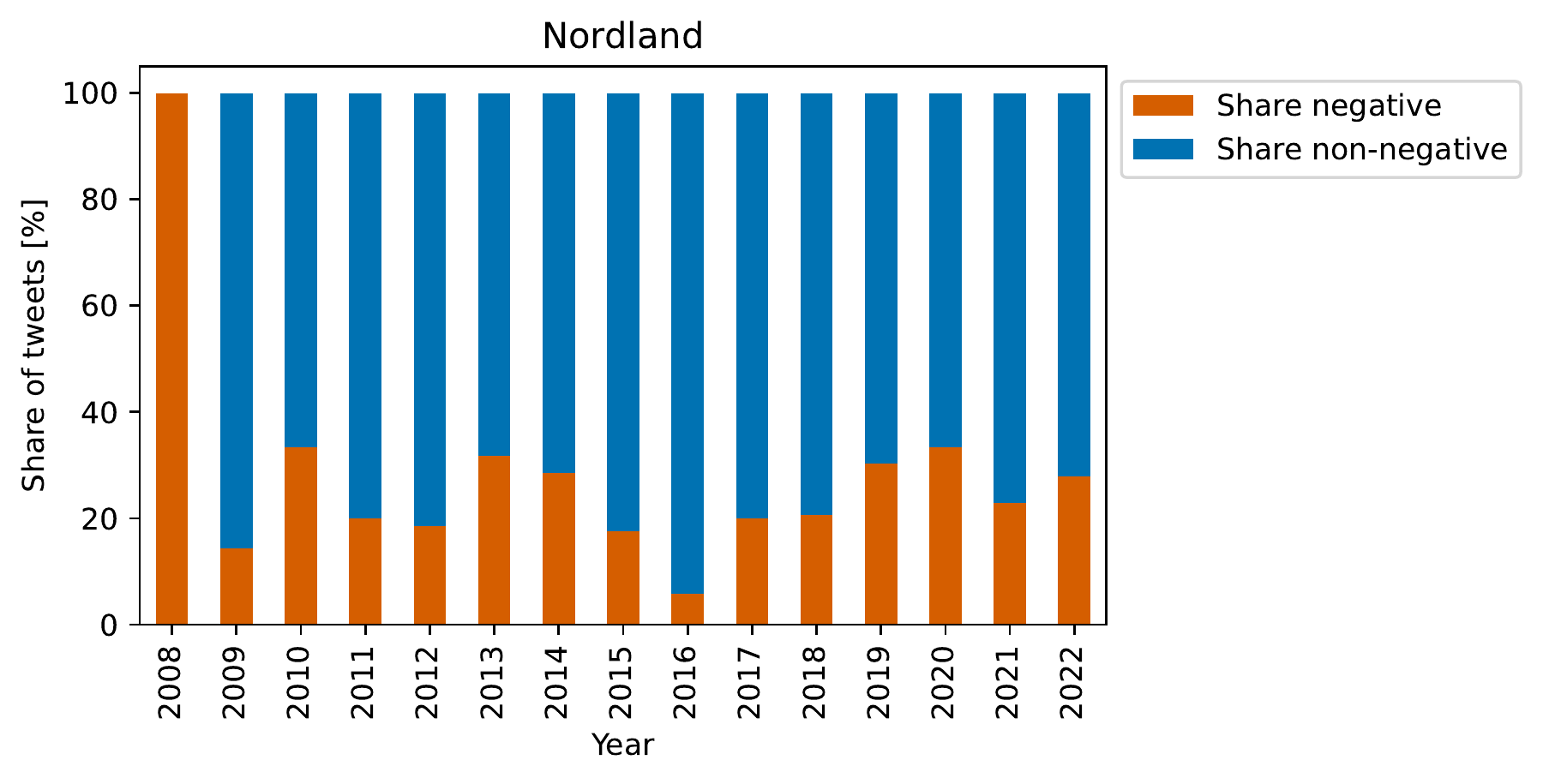}
    \caption{Share of positive and negative tweets per year for Nordland (N=915)}
    \label{fig:app_nord_bar}
\end{figure}

\newpage
\subsection{Oslo}
\begin{figure}[h!]
    \centering
    \includegraphics[width=0.8\textwidth]{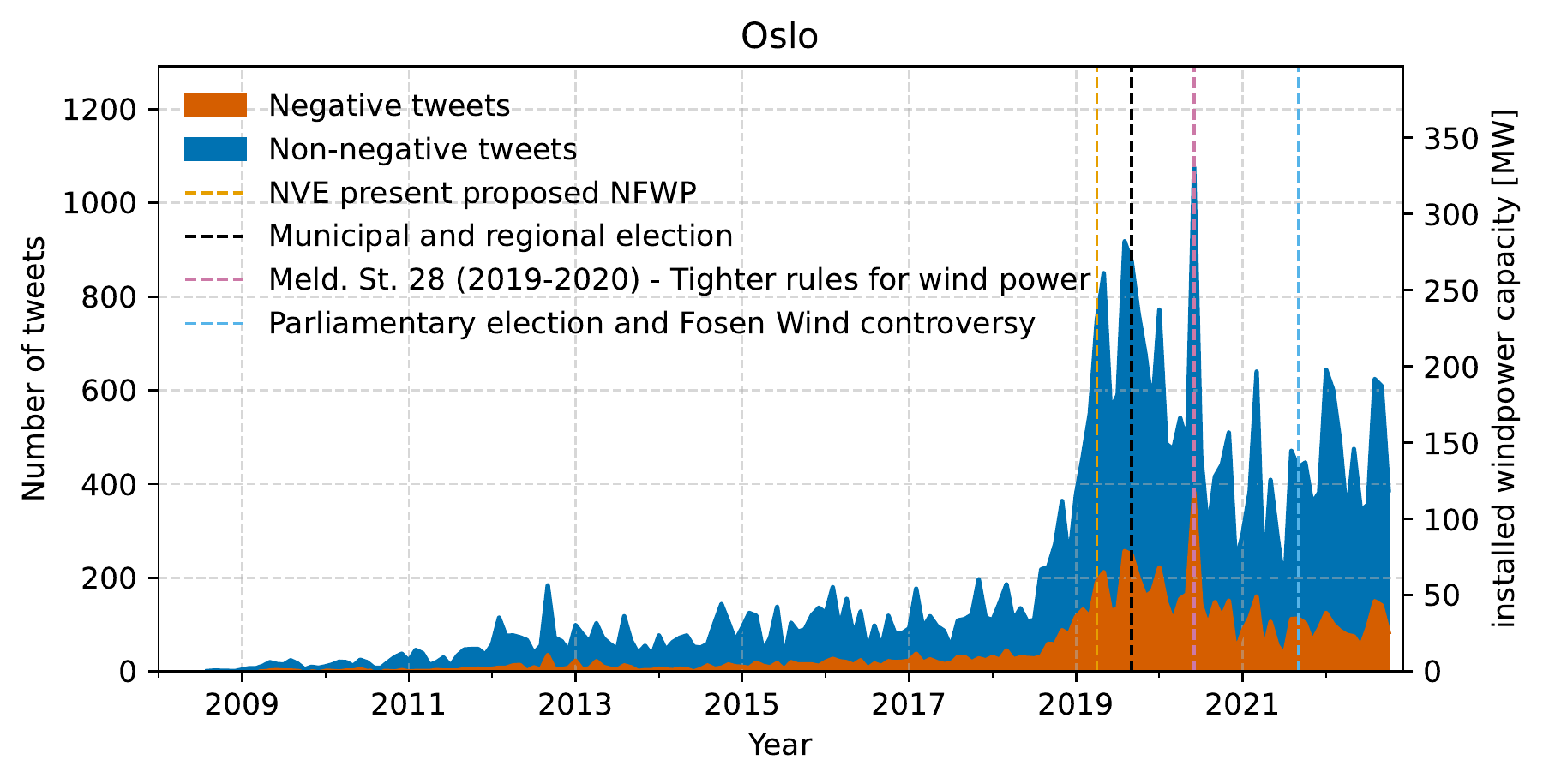}
    \caption{Temporal development of sentiment expressed on Twitter (monthly aggregation), installed wind power
capacity and possible influential events for Oslo}
    \label{fig:app_oslo_area}
\end{figure}

\begin{figure}[h!]
    \centering
    \includegraphics[width=0.8\textwidth]{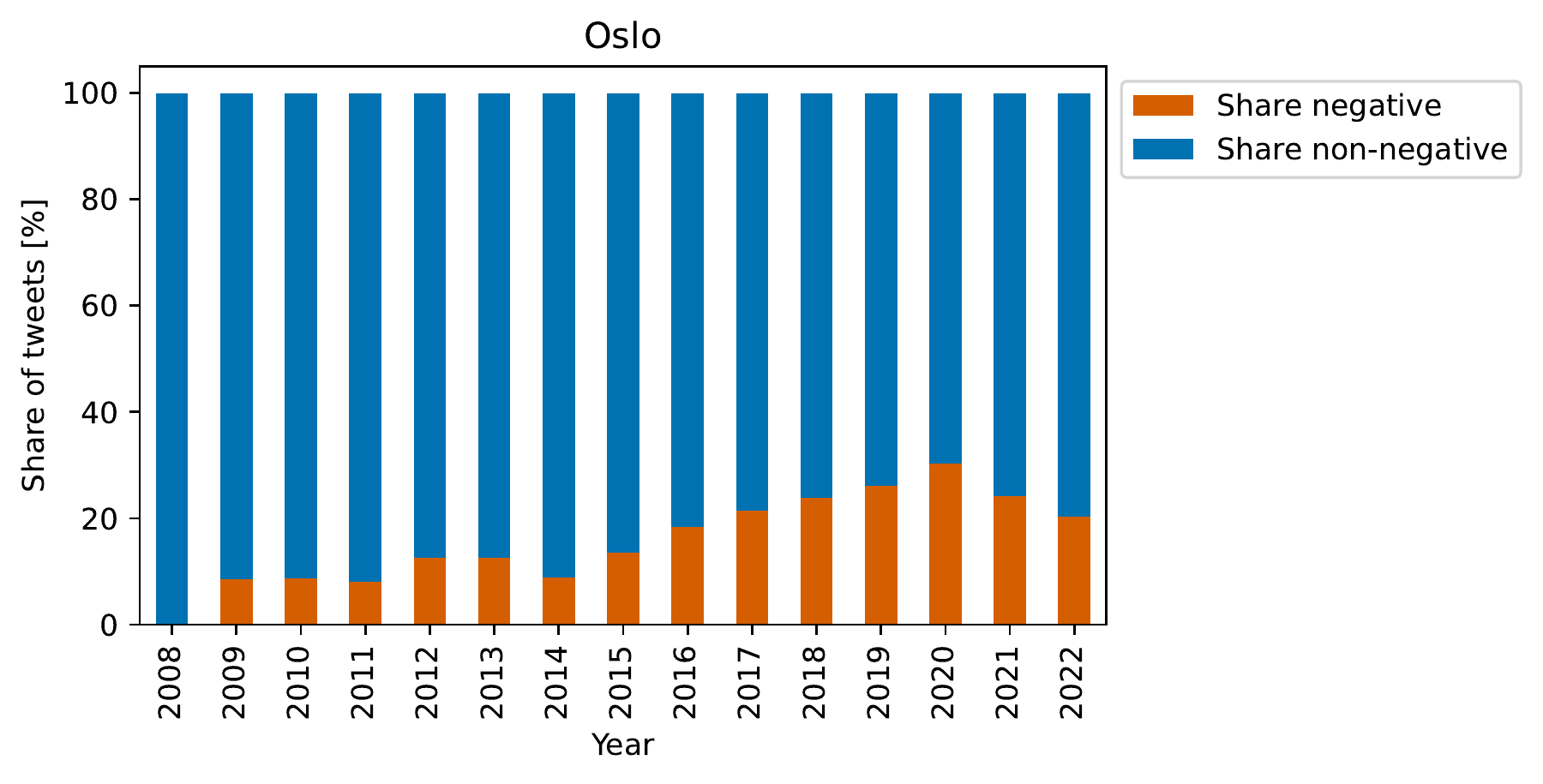}
    \caption{Share of positive and negative tweets per year for Oslo (N=32901)}
    \label{fig:app_oslo_bar}
\end{figure}

\newpage
\subsection{Rogaland}
\begin{figure}[h!]
    \centering
    \includegraphics[width=0.8\textwidth]{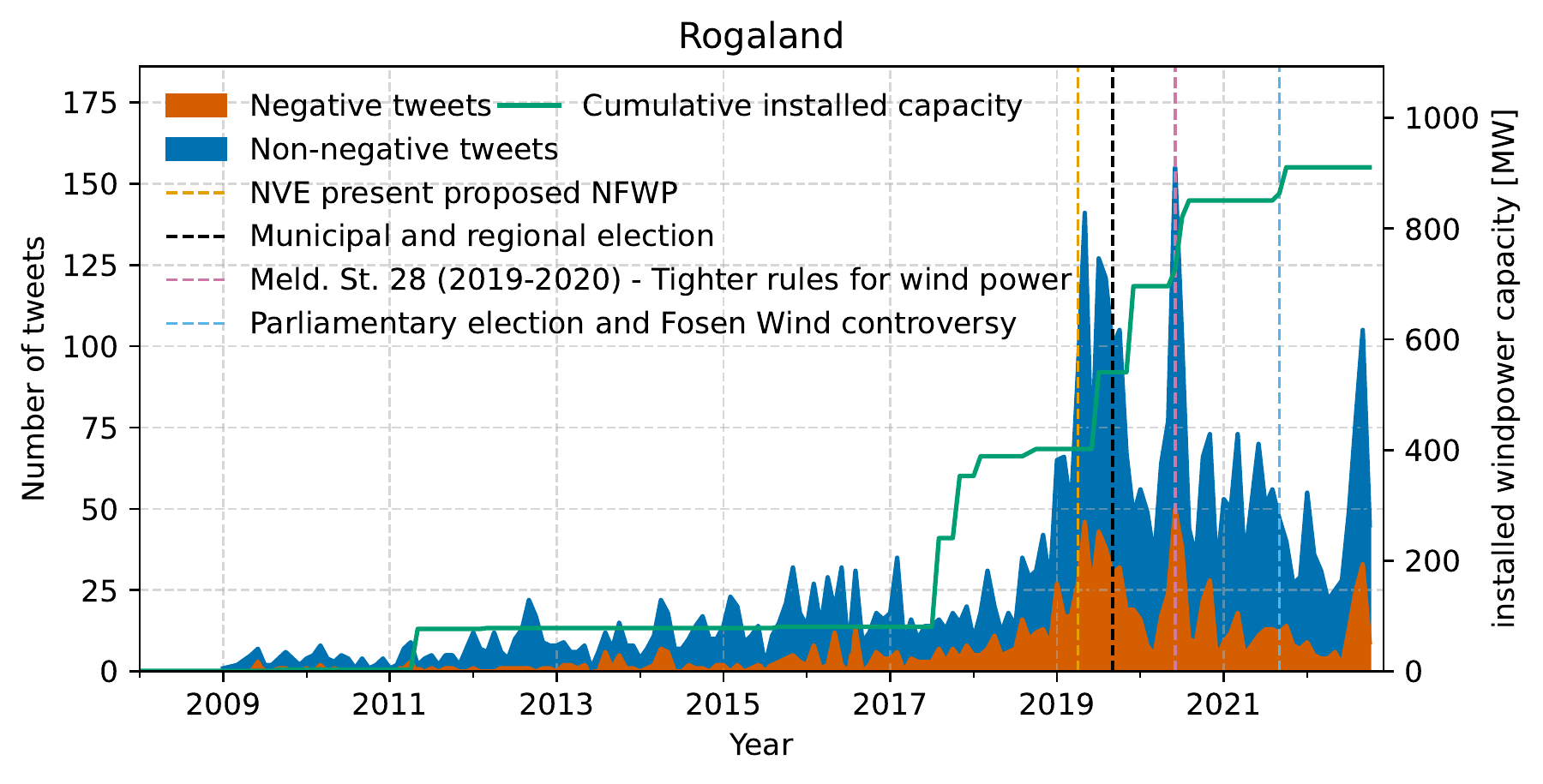}
    \caption{Temporal development of sentiment expressed on Twitter (monthly aggregation), installed wind power
capacity and possible influential events for Rogaland}
    \label{fig:app_rogaland_area}
\end{figure}

\begin{figure}[h!]
    \centering
    \includegraphics[width=0.8\textwidth]{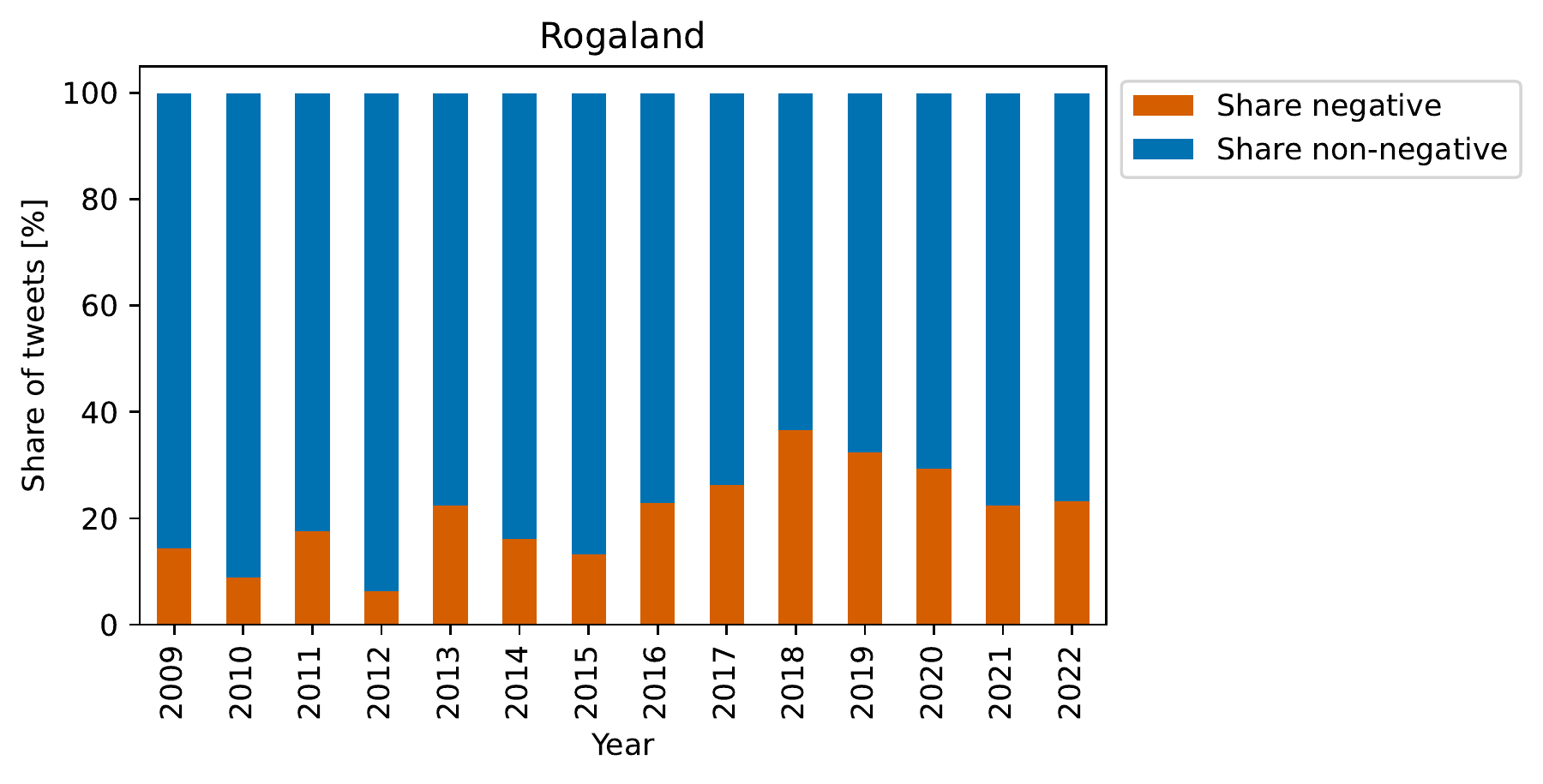}
    \caption{Share of positive and negative tweets per year for Rogaland (N=4280)}
    \label{fig:app_rogaland_bar}
\end{figure}

\newpage
\subsection{Troms og Finnmark}
\begin{figure}[h!]
    \centering
    \includegraphics[width=0.8\textwidth]{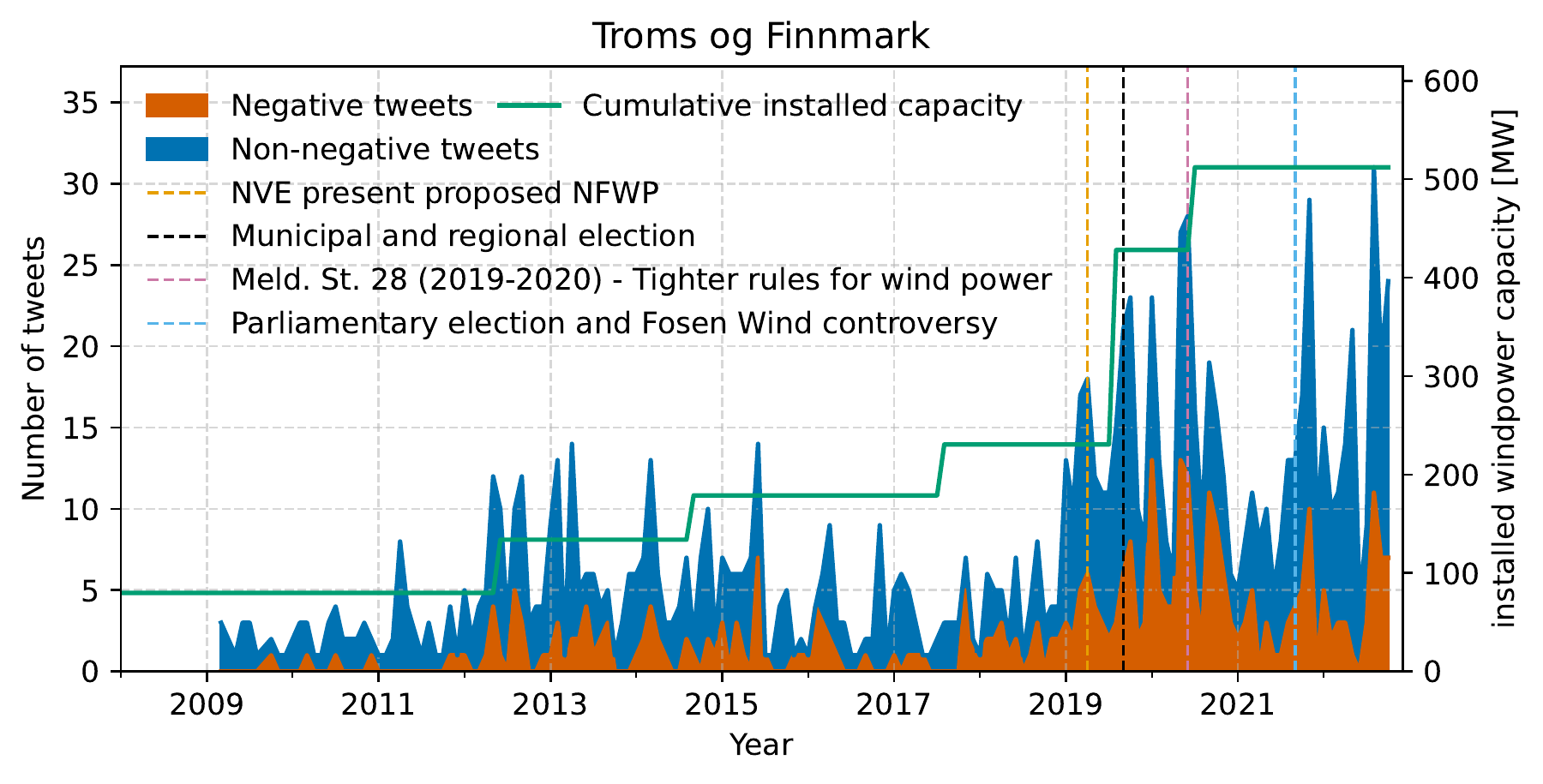}
    \caption{Temporal development of sentiment expressed on Twitter (monthly aggregation), installed wind power
capacity and possible influential events for Troms og Finnmark}
    \label{fig:app_troms_area}
\end{figure}

\begin{figure}[h!]
    \centering
    \includegraphics[width=0.8\textwidth]{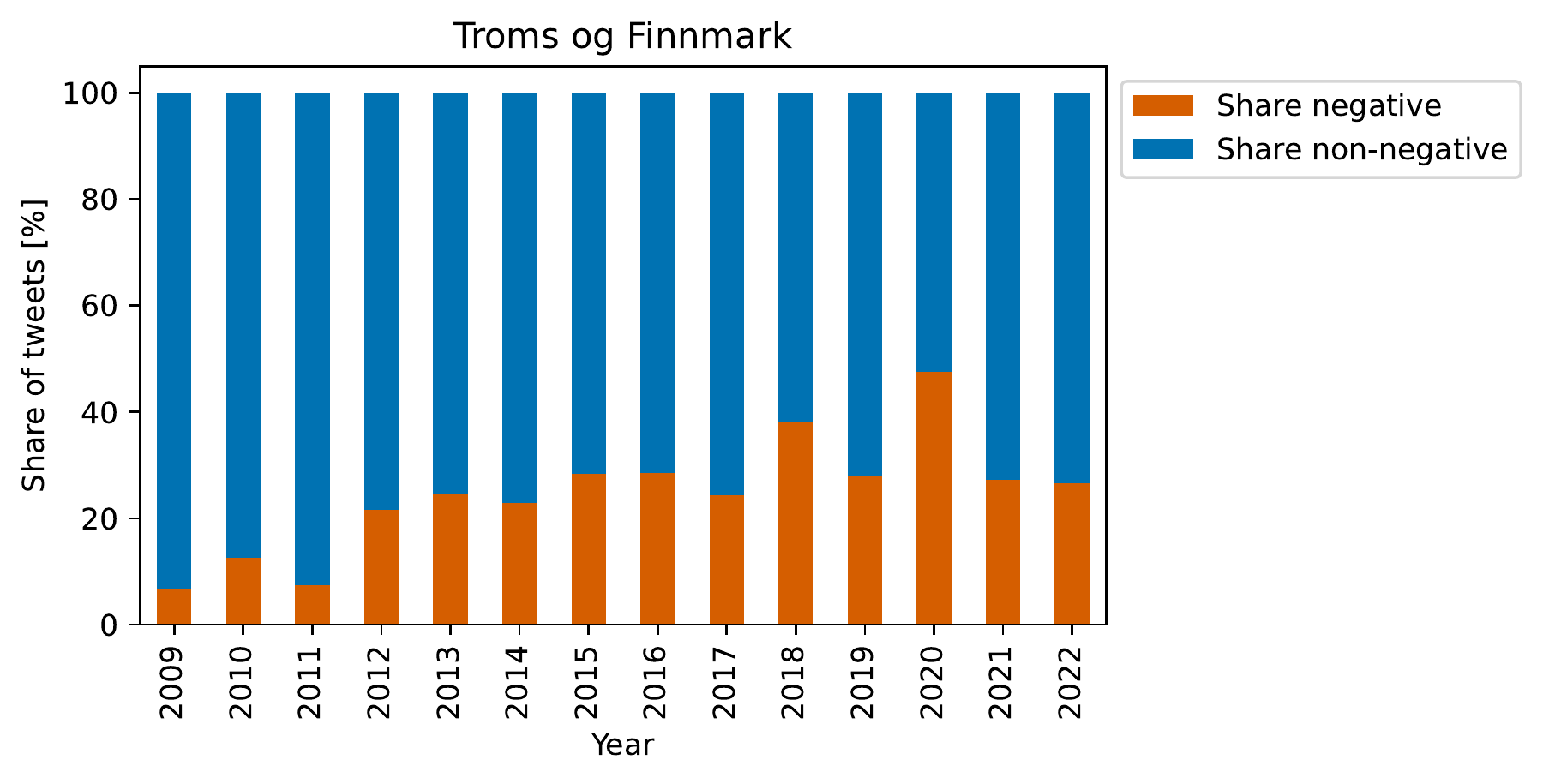}
    \caption{Share of positive and negative tweets per year for Troms og Finnmark (N=1122)}
    \label{fig:app_troms_bar}
\end{figure}

\newpage
\subsection{Trøndelag}
\begin{figure}[h!]
    \centering
    \includegraphics[width=0.8\textwidth]{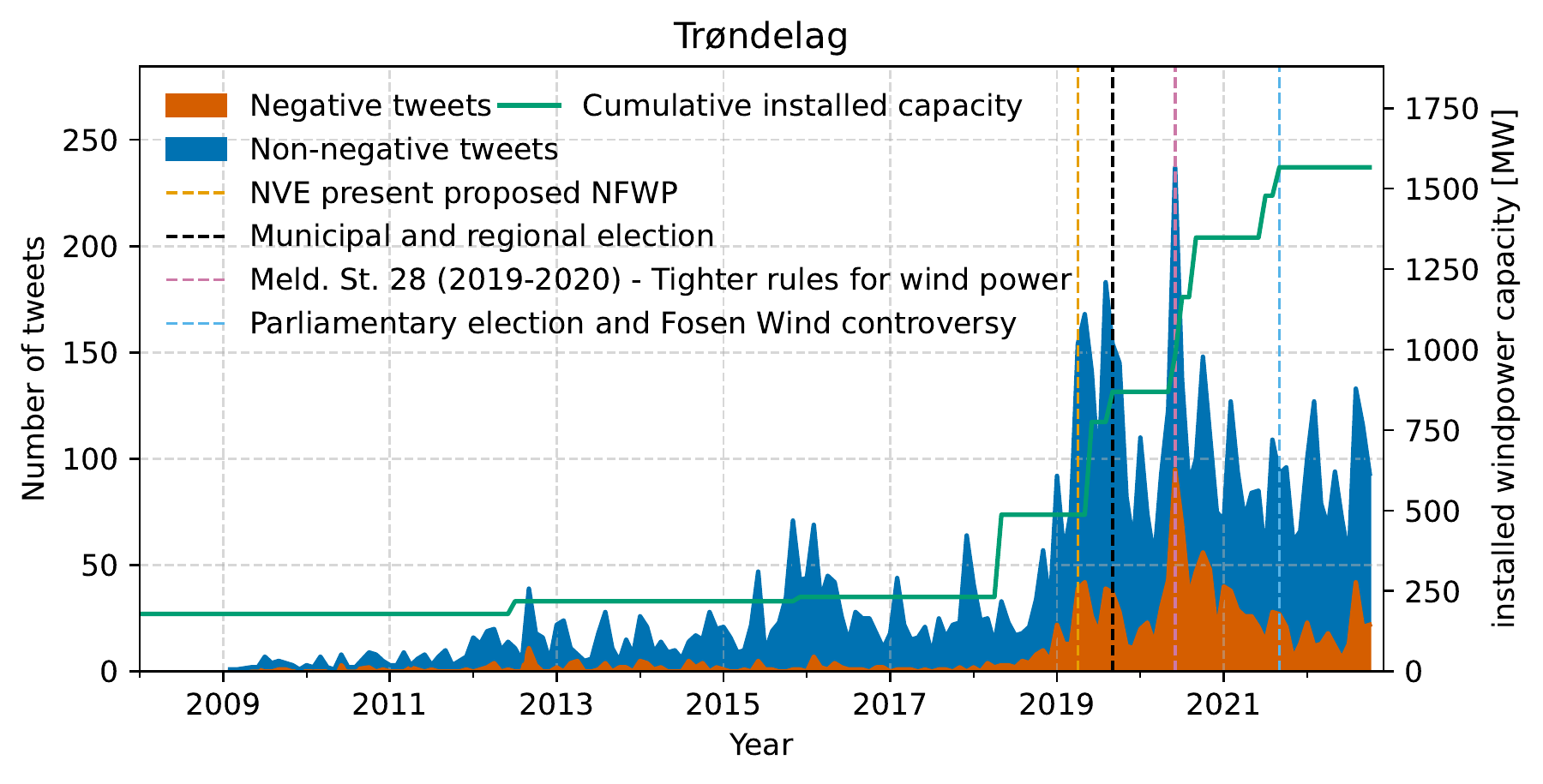}
    \caption{Temporal development of sentiment expressed on Twitter (monthly aggregation), installed wind power
capacity and possible influential events for Trøndelag}
    \label{fig:app_trond_area}
\end{figure}

\begin{figure}[h!]
    \centering
    \includegraphics[width=0.8\textwidth]{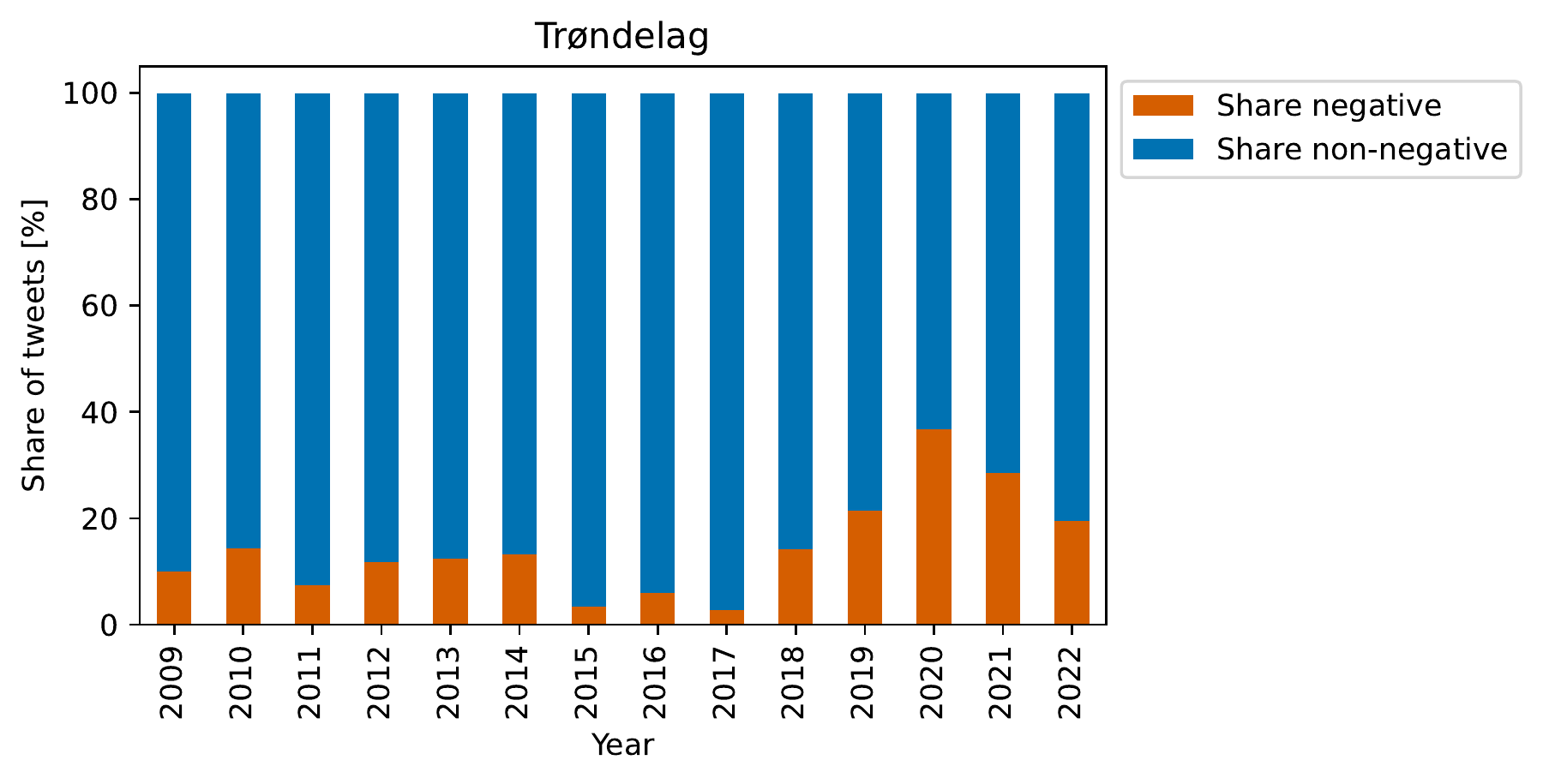}
    \caption{Share of positive and negative tweets per year for Trøndelag (N=6718)}
    \label{fig:app_trond_bar}
\end{figure}

\newpage
\subsection{Vestfold og Telemark}
\begin{figure}[h!]
    \centering
    \includegraphics[width=0.8\textwidth]{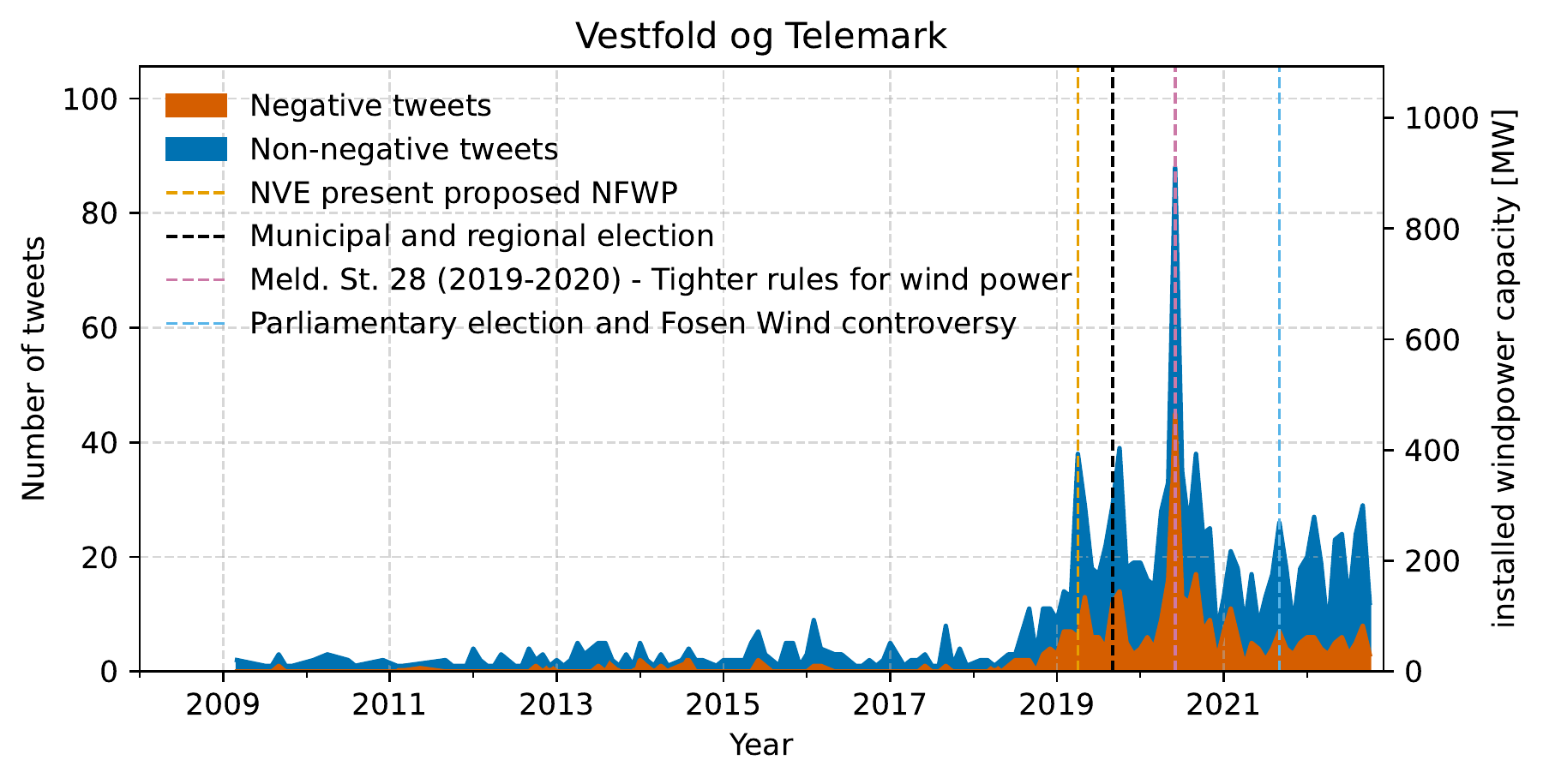}
    \caption{Temporal development of sentiment expressed on Twitter (monthly aggregation), installed wind power
capacity and possible influential events for Vestfold og Telemark}
    \label{fig:app_vestfold_area}
\end{figure}

\begin{figure}[h!]
    \centering
    \includegraphics[width=0.8\textwidth]{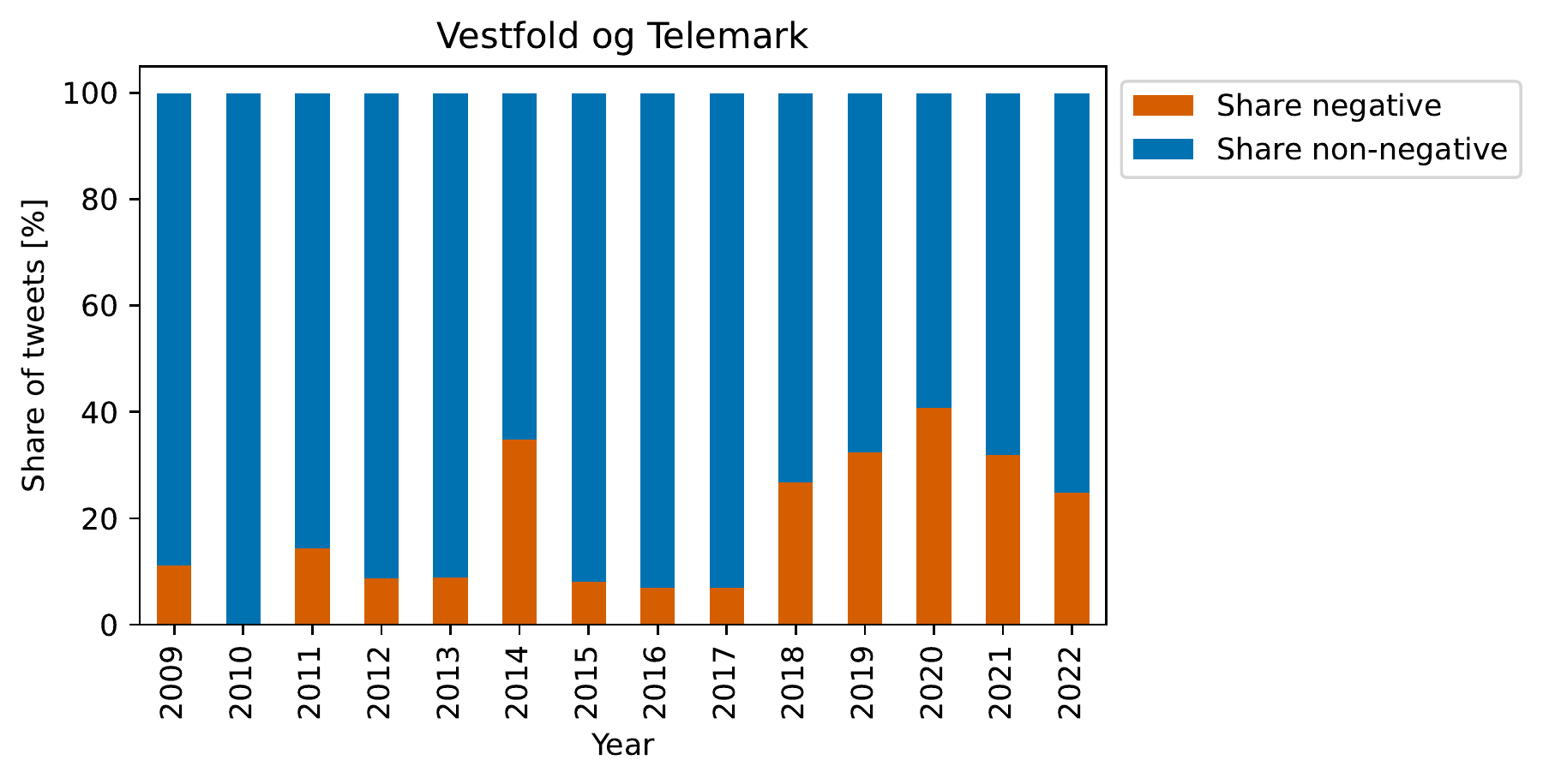}
    \caption{Share of positive and negative tweets per year for Vestfold og Telemark (N=1257)}
    \label{fig:app_vestfold_bar}
\end{figure}

\newpage
\subsection{Vestland}
\begin{figure}[h!]
    \centering
    \includegraphics[width=0.8\textwidth]{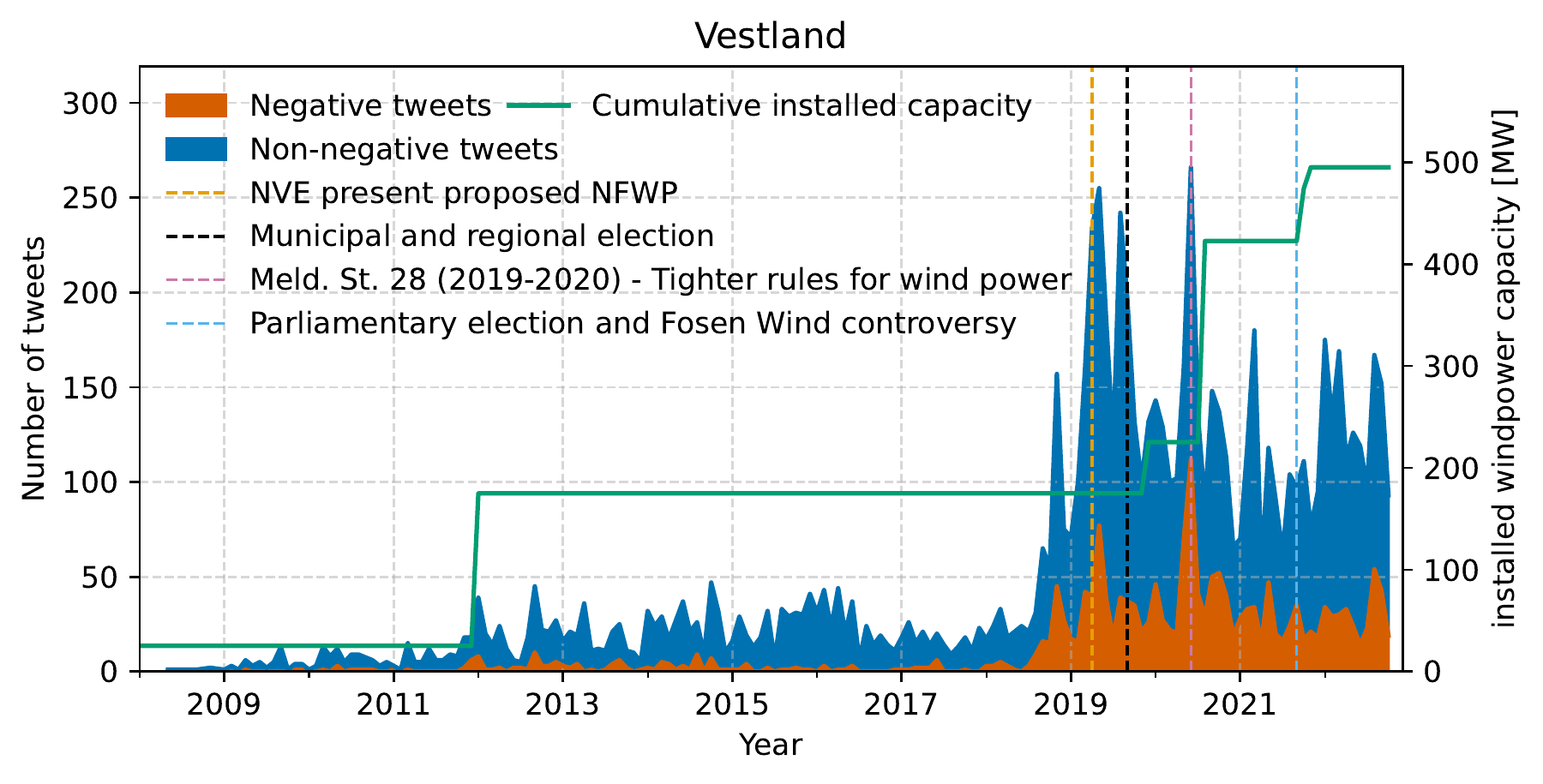}
    \caption{Temporal development of sentiment expressed on Twitter (monthly aggregation), installed wind power
capacity and possible influential events for Vestland}
    \label{fig:app_vestland_area}
\end{figure}

\begin{figure}[h!]
    \centering
    \includegraphics[width=0.8\textwidth]{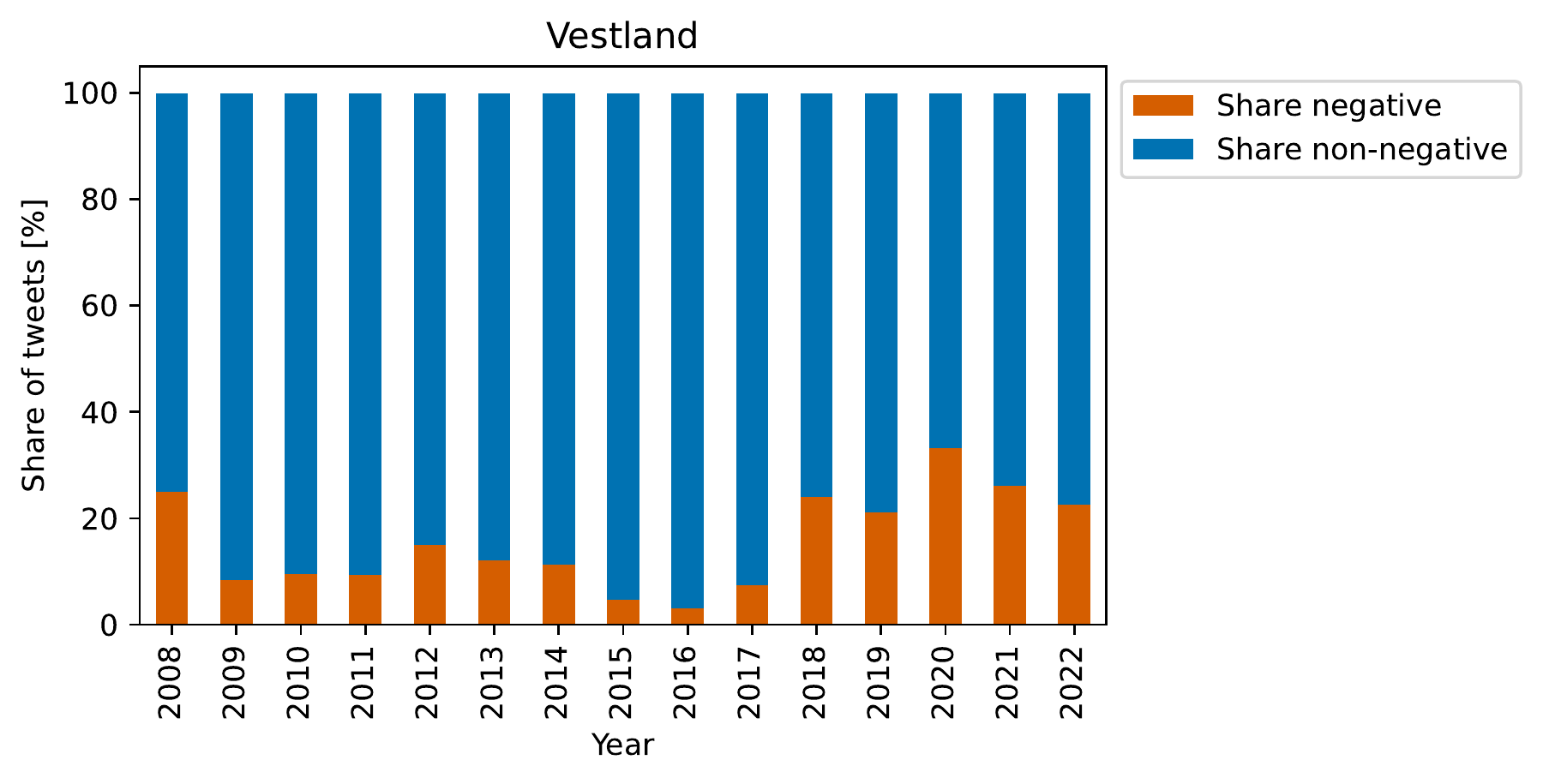}
    \caption{Share of positive and negative tweets per year for Vestland (N=8354)}
    \label{fig:app_vestland_bar}
\end{figure}

\newpage
\subsection{Viken}
\begin{figure}[h!]
    \centering
    \includegraphics[width=0.8\textwidth]{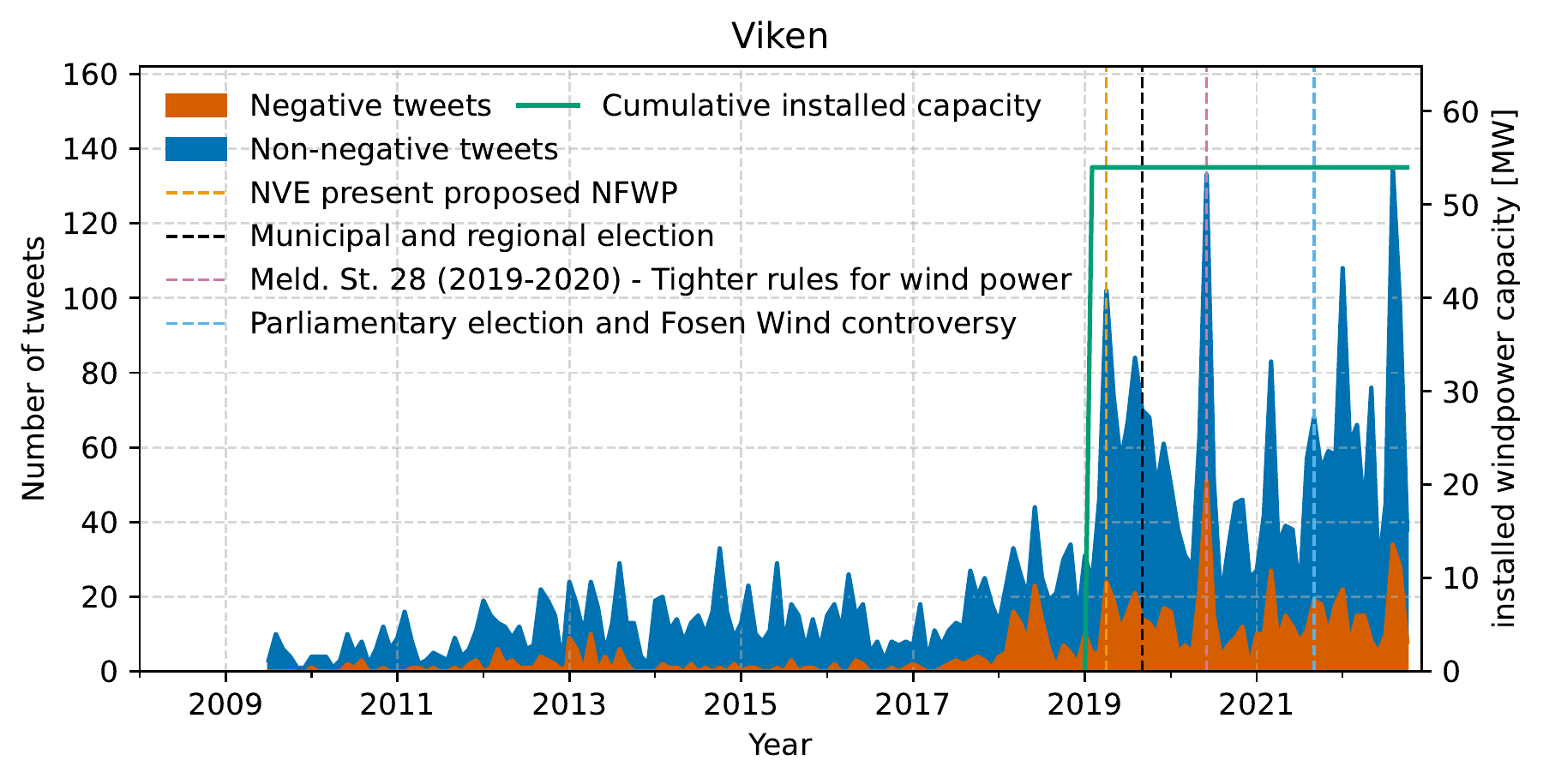}
    \caption{Temporal development of sentiment expressed on Twitter (monthly aggregation), installed wind power
capacity and possible influential events for Viken}
    \label{fig:app_viken_area}
\end{figure}

\begin{figure}[h!]
    \centering
    \includegraphics[width=0.8\textwidth]{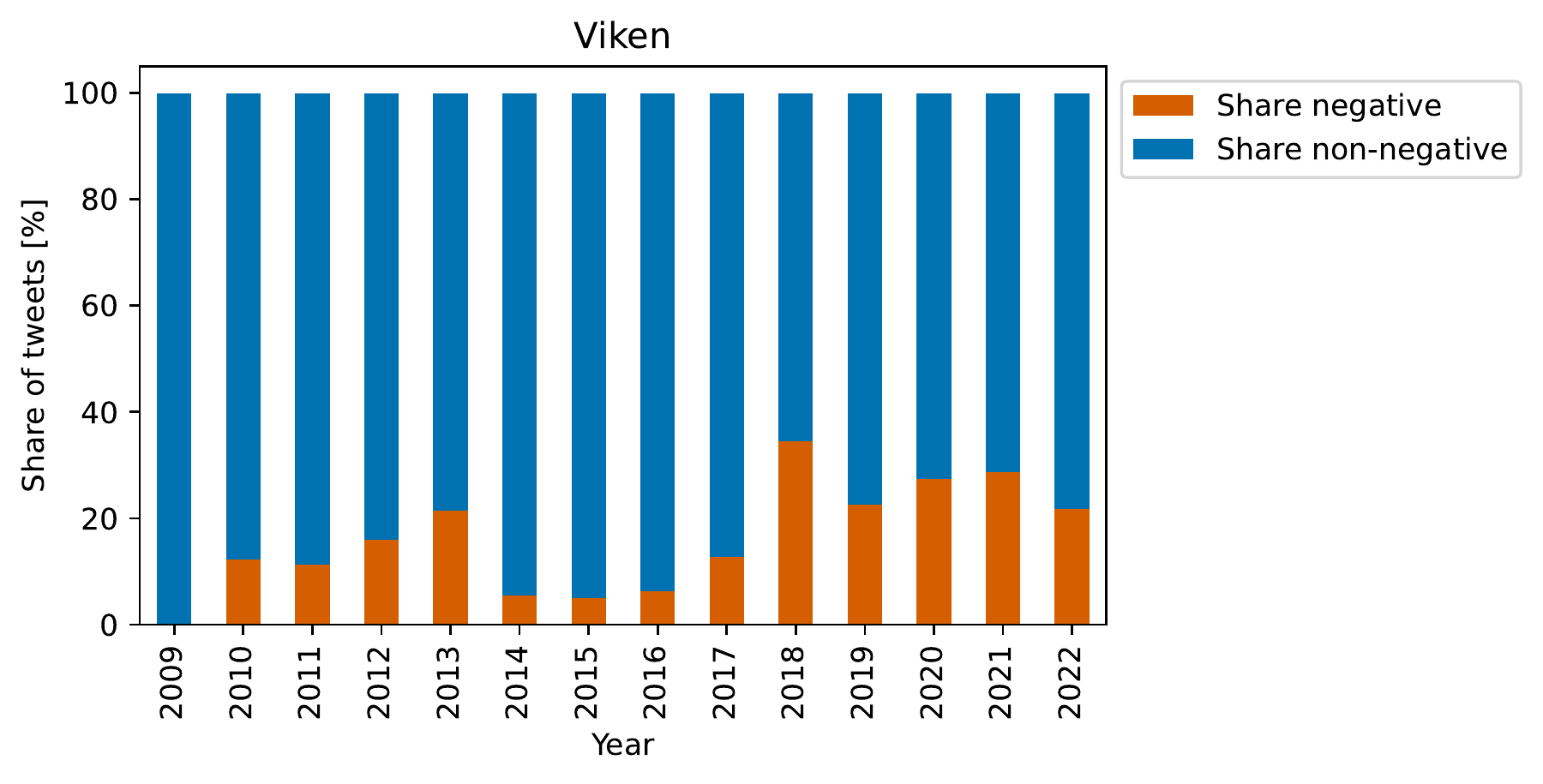}
    \caption{Share of positive and negative tweets per year for Viken (N=4023)}
    \label{fig:app_viken_bar}
\end{figure}

\newpage
\section{Heat maps of share of negative sentiment between 2019-2022}
\label{app:heat_maps}

\begin{figure}[h!]
    \centering
    \includegraphics[width=\textwidth]{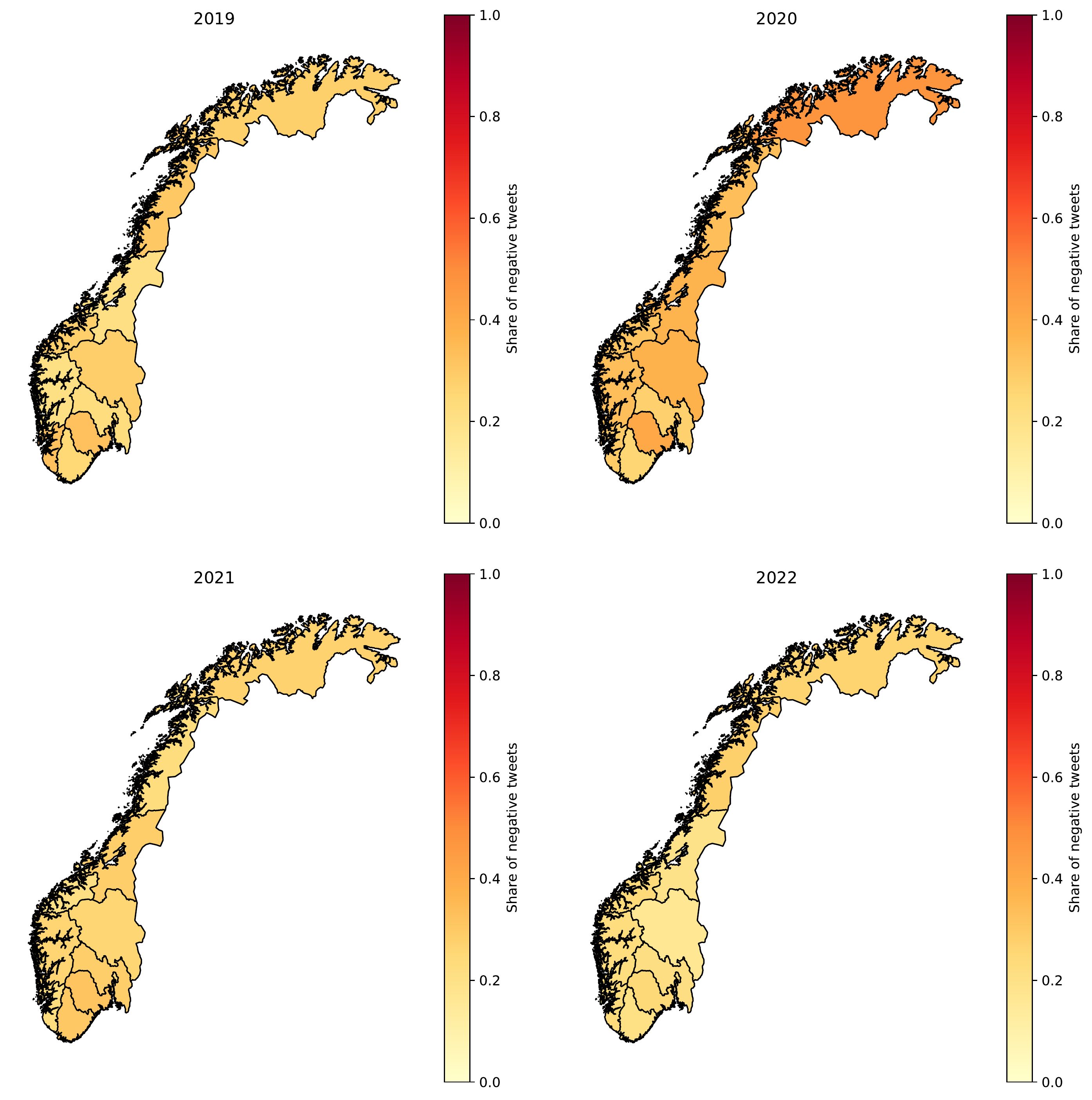}
    \caption{Heatmap of the share of negatively categories tweets for each of the Norwegian counties (NUTS3 level) for 2019-2022}
    \label{fig:county_sentiment}
\end{figure}

\newpage
\section{Normalised regional trends}
\label{app:regional_trends}

\begin{figure}[h!]
    \centering
    \includegraphics[width=\textwidth]{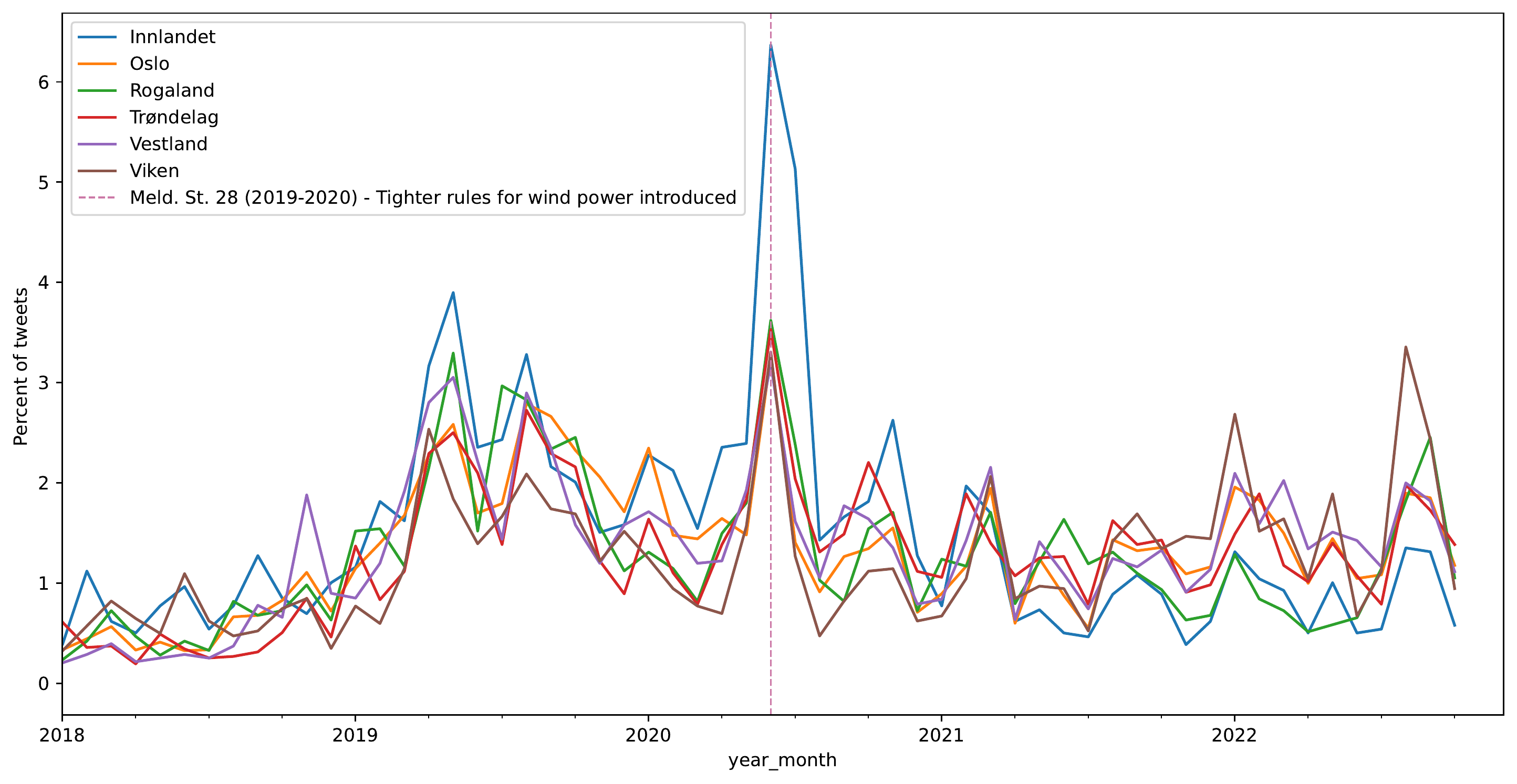}
    \caption{Normalised trends of tweets between 2018 and 2022.}
    \label{fig:app_relative_regional_trends}
\end{figure}

\newpage
\section{Twitter networks}
\label{app:network_analysis}

\begin{figure}[h!]
    \centering
    \includegraphics[width=\textwidth]{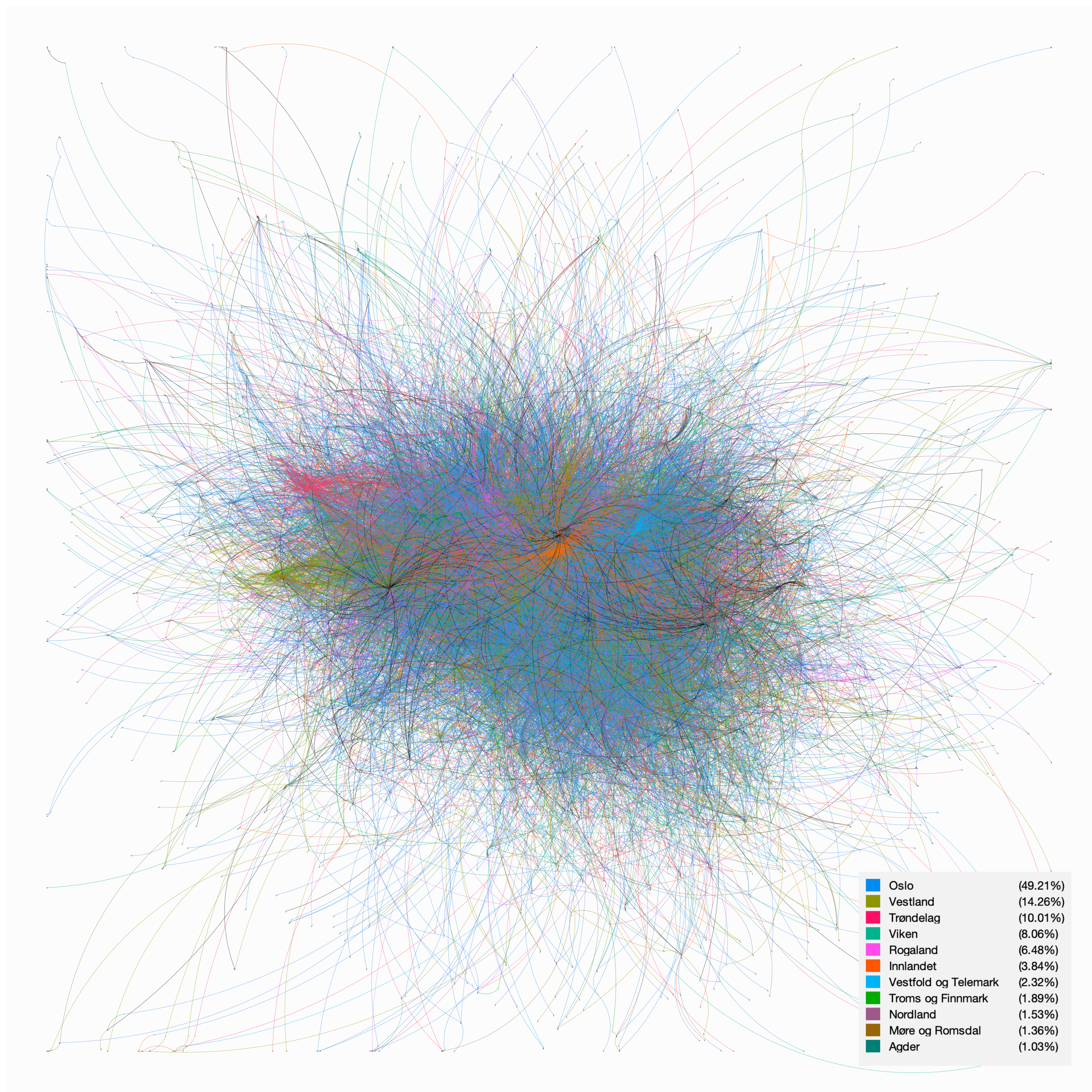}
    \caption{Network graph. Edges are colored according to the county of the user who sent the tweet. The network has been visualized using the ForceAtlas2 algorithm in Gephi \citep{jacomyForceAtlas2ContinuousGraph2014}. In order to see if users clustered according to region, users were first grouped into community clusters using the modularity algorithm \citep{blondelFastUnfoldingCommunities2008}; a chi-square test of independence was then performed between users' counties and their community clusters, $\chi^2$ (50, N = 60450) = 6092.78, p < .001, V = .142. }
    \label{fig:network_graph}
\end{figure}

\end{document}